\documentclass[a4paper]{aa}
\usepackage{graphics,txfonts,amsfonts}

\usepackage{natbib}
\bibpunct[, ]{(}{)}{;}{a}{}{,}

\newcommand{\teff}{$T_{\rm eff}$}
\newcommand{\lgg}{$\log g$}
\newcommand{\vs}{$v_{\rm e}\sin i$}

\newcommand{\bz}{$\langle B_z\rangle$}
\newcommand{\kms}{km\,s$^{-1}$}

\newcommand{\cvn}{$\alpha^2$\,CVn}
\newcommand{\cam}{53\,Cam}
\newcommand{\inv}{{\sc Invers10}}
\newcommand{\mus}{{\sc MuSiCoS}}

\newcommand{\bb}{\mbox{{\boldmath $B$}}}

\newcommand{\figps}[1]{\resizebox{\hsize}{!}{\rotatebox{0}{\includegraphics{#1}}}}

\newcommand{\fifps}[2]{\centering\resizebox{#1}{!}{\includegraphics{#2}}}

\newcommand{\figgps}[2]{\resizebox{#1}{!}{\rotatebox{0}{\includegraphics{#2}}}}

\newcommand{\beq}{\begin{equation}}
\newcommand{\eeq}{\end{equation}}

\begin{document}

\title{Magnetic Doppler imaging of $\alpha^2$\,Canum Venaticorum in \\ all four Stokes parameters%
\thanks{Based on data obtained using the T\'elescope Bernard Lyot at Observatoire du Pic du Midi}}

\subtitle{Unveiling the hidden complexity of stellar magnetic fields}

\author{O. Kochukhov\inst{1}
   \and G. A. Wade\inst{2}}

\offprints{O. Kochukhov, \\ \email{oleg.kochukhov@fysast.uu.se}}

\institute{Department of Physics and Astronomy, Uppsala University, Box 516, Uppsala SE-751 20, Sweden
      \and Department of Physics, Royal Military College of Canada, Box 17000, Kingston, Ontario, K7K 4B4 Canada}

\date{Received 14 December 2009 / Accepted 28 January 2010}

\abstract%
% Context
{Strong organized magnetic fields have been studied in the upper main sequence chemically peculiar stars for more than half a century. However, only recently have observational methods and numerical techniques 
become sufficiently mature to allow us to record and interpret high-resolution four Stokes parameter spectra, leading to the first assumption-free magnetic field models of these stars. 
}
% Aims
{
Here we present a detailed magnetic Doppler imaging analysis of the spectropolarimetric observations of the prototypical magnetic Ap star \cvn. This is the second star for which the magnetic field topology and horizontal chemical abundance inhomogeneities have been inferred directly from phase-resolved observations of line profiles in all four Stokes parameters, free from the traditional assumption of a low-order multipolar field geometry.
}
% Method
{
We interpret the rotational modulation of the circular and linear polarization profiles of the strong \ion{Fe}{ii} and \ion{Cr}{ii} lines in the spectra of \cvn\ recorded with the \mus\ spectropolarimeter. The surface abundance distributions of the two chemical elements and a full vector map of the stellar magnetic field are reconstructed in a self-consistent inversion using our state-of-the-art magnetic Doppler imaging code \inv. 
%based on realistic theoretical calculation of the polarized line profiles of individual metal lines.
}
% Results
{
We succeeded in reproducing most of the details of the available spectropolarimetric observations of \cvn\ with a magnetic map which combines a global dipolar-like field topology with localized spots of higher field intensity. We demonstrate that these small-scale magnetic structures are inevitably required to fit the linear polarization spectra; however, their presence cannot be inferred from the Stokes $I$ and $V$ observations alone. We also found high-contrast surface distributions of Fe and Cr, with both elements showing abundance minima in the region of weaker and topologically simpler magnetic field. 
}
% Conclusions
{
Our magnetic Doppler imaging analysis of \cvn\ and previous results for \cam\ support the view that the upper main sequence stars can harbour fairly complex surface magnetic fields which resemble oblique dipoles only at the largest spatial scales. Spectra in all four Stokes parameters are absolutely essential to unveil and meaningfully characterize this field complexity in Ap stars. We therefore suggest that understanding magnetism of stars in other parts of the H-R diagram is similarly incomplete without investigation of their linear polarization spectra.
}

\maketitle

\section{Introduction}
\label{intro}

Magnetic fields play a fundamental role in the physics of the atmospheres of a significant fraction of stars on the H-R diagram. The magnetic fields of intermediate-mass upper main sequence stars (the Ap stars) have very different characteristics, and probably a different origin, than those of late- type stars like the sun \citep[e.g.][]{mestel:2003}. In Ap stars, the large-scale surface magnetic field is observed to be static on timescales of at least many decades, and appears to be ``frozen'' into a rigidly rotating atmosphere. The magnetic field is globally organized, permeating the entire stellar surface, with a relatively high field strength (typically of a few hundreds up to a few tens of thousands of gauss). Magnetic field appears to be present in only a small fraction of intermediate-mass main sequence stars, and its presence strongly influences energy and mass transport (e.g., diffusion, convection and weak stellar winds) within the atmosphere. The characteristic consequence of this interaction is the presence of strong chemical abundance non-uniformities in photospheric layers. 

The weight of opinion holds that the magnetic fields of upper main sequence stars are primarily fossil fields -- the slowly-decaying remnants of magnetic field accumulated or generated during the complex process of star formation. In this framework, the global and local properties of the magnetic field (the mean intensity, obliquity relative to the stellar rotation axis, and large-scale topology; but also its smaller-scale structure) may well provide unique information about invisible interior processes which have occurred or are occurring within the star: differential rotation, meridional circulation currents, global- and local-scale dynamo action, etc. Therefore a knowledge of the detailed magnetic field structure of Ap stars -- on both large and small scales -- can contribute significantly to understanding the physics of such stars.

The characteristics of magnetic fields in upper main sequence stars are inferred from the influence of the Zeeman effect on their spectra. Even a relatively weak magnetic field (a few tens of gauss) will clearly imprint its presence on the local emergent spectrum of a star, by splitting and polarizing line profiles. The large majority of magnetic field data in the literature measure the longitudinal Zeeman effect via induced circular polarization within photospheric absorption lines, as quantified by the mean longitudinal magnetic field. While such measurements represent a powerful tool for detecting organized magnetic fields, they are relatively insensitive to the topology of the magnetic field, typically constraining the strength and orientation of the field's dipole component. If line circular polarization (the $V$ Stokes parameter) is measured using high spectral resolving power, additional detail can be recovered by exploiting the rotational Doppler effect via Magnetic Doppler Imaging \citep[MDI; see][]{piskunov:2002a,kochukhov:2002c}. This was demonstrated by \citet{kochukhov:2002b}, who mapped the magnetic field of the Ap star \cvn\ using a timeseries of high-resolution Stokes $I$ and $V$ line profiles, interpreted using MDI. To converge to a unique solution, the map of \cvn\ was forced to resemble, to the greatest extent possible given the constraints imposed by the data, a non-axisymmetric multipolar configuration. The resulting map showed a dominant dipole component, with a small quadrupolar contribution. Overall, the map was quite smooth, in agreement with expectations.

Additional magnetic field structural detail is accessible through the spectral line linear polarization ($Q$ and $U$ Stokes parameters) induced by the transverse Zeeman effect. As demonstrated by \citet{leroy:1995a}, Zeeman linear polarization, even when measured photometrically using broad bandpasses, provides information about the smaller-scale structure of the magnetic field that is not available from circular polarization. \citet{wade:2000b} extended these observational investigations to their natural culmination by obtaining high-precision measurements of Zeeman polarization in both circular and linear polarization (i.e. in all four Stokes $I$, $V$, $Q$ and $U$ parameters) for a sample of about 15 Ap stars, with a resolving power sufficient to distinguish the variation of the polarization across individual spectral lines. For about a dozen of these stars, multiple Stokes $IQUV$ sequences were acquired, allowing a characterisation of the longitudinal and transverse components of the magnetic field on different regions of the stellar surface. For the A2p star 53~Cam, \citet{wade:2000b} and \citet{bagnulo:2001} demonstrated that the existing low-order multipolar models of the magnetic field were unable to reproduce the observed Stokes $Q$ and $U$ spectra, although they acceptably matched the Stokes $V$ spectra. They interpreted their results as indicating that the magnetic topology of 53~Cam was significantly more complex than that envisioned by the models.

To test this proposal, \citet{kochukhov:2004c} applied MDI to the 53~Cam data of \citet{wade:2000b} and \citet{bagnulo:2001} -- the first attempt to reconstruct a stellar magnetic field using measurements of spectral lines in the four Stokes parameters. Kochukhov et al. were able to reproduce the variable intensity and morphology of the polarized line profiles of 3 strong Fe~{\sc ii} lines throughout the star's rotational cycle. In agreement with the prediction of \citet{wade:2000b} and \citet{bagnulo:2001}, the derived field topology was surprisingly complex. In particular, while the radial field exhibited an approximately dipolar behaviour, the tangential field and the field modulus showed structure on much smaller scales. 

In the context of their analysis of 53~Cam, Kochukhov et al. demonstrated that the constraint on the transverse field provided by linear polarization measurements was essential to detecting the fine structure of the field. It was therefore natural to revisit the analysis of \cvn, to examine to what extent the smooth map derived by \citet{kochukhov:2002b} is attributable to their lack of linear polarization data. This is the topic of the present paper. In Sect.~\ref{obs} we describe the spectropolarimetric observations from which the MDI maps are derived. In Sect.~\ref{props} we determine the physical parameters of the star, with the particular goal of constraining the 
%inclination 
orientation of the stellar rotation axis with respect to the line-of-sight. In Sect.~\ref{mdi} we review basic principles of MDI, and describe the spectral lines used, their associated atomic data, and the determination of global parameters to be used in the mapping procedure. In Sect.~\ref{results} we describe the results of the mapping, paying particular attention to the significance and robustness of small-scale structures detected in the field modulus map. 

\section{Spectropolarimetric observations}
\label{obs}

The main set of spectropolarimetric observations of \cvn\ was obtained in 1997--1999 using the (now decommissioned) \mus\ spectropolarimeter at the 2-m Bernard Lyot Telescope (TBL) at Pic du Midi observatory.

The \mus\ instrument consists of a table-top cross-dispersed \'echelle spectrograph  \citep{baudrand:1992}, fed by optical fibres directly from a Cassegrain-mounted polarization analysis module. This instrument allows the acquisition of a stellar spectrum in a given polarization state (Stokes $V$, $Q$ or $U$) throughout the spectral range 450 to 660~nm in a single exposure. The resolving power is $\lambda/\Delta\lambda\simeq$ \,35\,000. The optical characteristics of the spectropolarimeter and corresponding observing procedures are described in detail by \citet{donati:1999a}.

When \mus\ is used for polarization observations, starlight enters the polarimeter at the Cassegrain focus. The beam then may optionally pass through a rotatable $\lambda/4$ retarder (in the case of Stokes $V$ observations) or not (in the case of Stokes $Q$ or $U$ observations). The beam then intersects a Savart-type beamsplitter which separates the stellar light into two beams which are respectively polarized along and perpendicular to the instrumental reference azimuth. The analyzed beams are then injected into the double 50~$\mu$m fibre, which transports the light to the spectrograph. Spectra in both orthogonal polarizations are thereby recorded simultaneously on the thinned $1024\times 1024$ pixel {\sc site} CCD detector.

A single polarimetric observation, yielding an intensity spectrum and one other Stokes parameter, consists of a sequence of 4 subexposures, between which the retarder (for circular polarization Stokes $V$) or the polarimetric module itself (for linear polarizations Stokes $Q$ and $U$) is rotated by $\pm90\degr$ following the procedure suggested by \citet{semel:1993}. This has the effect of exchanging the beams within the whole instrument and switching the positions of the two orthogonally polarized spectra on the CCD. This observing procedure suppresses all first-order spurious polarization signatures down to well below the noise level.

Spectra obtained at the TBL using the \mus\ spectrograph and polarimeter were reduced using the {\sc ESpRIT} reduction package \citep{donati:1997}. The {\sc ESpRIT}-reduced spectra were then post-processed to improve the quality of the continuum normalization. A full description of the acquisition and reduction procedure of the \mus\ observations of \cvn\ is provided by \citet{wade:2000b}.

A complete list of the spectropolarimetric observations of \cvn\ analyzed in our study is presented in Table~\ref{tbl:obs}. This table provides information on the UT date of observation, lists Stokes parameters obtained, their Julian dates, mean rotation phases and the signal-to-noise ratio. The Stokes $V$ spectrum recorded on 25 February 1997 was treated separately from the linear polarization observations on that night due to a gap in observing time exceeding 1\% of the rotation period. 

A close examination of the Stokes $Q$ and $U$ profiles obtained on 20 February 1997 ($\varphi=0.391$, see \citealt{wade:2000b}), comparing with the observations closest in phase, suggests that their signs are inverted. These particular linear polarization spectra were obtained with an interruption in the observing procedure, which possibly led to a sign error. We have obtained an additional observation of \cvn\ on 26 July 2004, at the rotation phase $\varphi=0.413$. The Stokes $Q$ and $U$ profiles from this phase are inverted in comparison to the Feb 20, 1997 observation and agree with the other spectra obtained in 1997--1999. Consequently, we decided to disregard the linear polarization spectra from 20 February 1997, using only the Stokes $V$ data from that night. 

Thus, in the magnetic DI analysis presented below we have modeled 20 separate phases of partial or complete Stokes parameter observations of \cvn. For the rotation phases where 2 or 3 intensity profiles were available with each of the other Stokes parameters, we have constructed an average Stokes $I$ spectrum. All spectra were then interpolated onto a common wavelength grid as required for comparison with theoretical calculations.

All observations of \cvn\ are phased using the ephemeris of \citet{farnsworth:1932}:
\beq
JD (Eu_{\mathrm{max}})=2419869.720 + 5\fd46939\times E,
\eeq
which gives the time of the maximum intensity of the spectral lines of \ion{Eu}{ii} and roughly corresponds to the negative extremum of the longitudinal magnetic field phase curve \citep{wade:2000}.

\begin{table*}[!th]
%\begin{center}
\caption{Four Stokes parameter
spectropolarimetric observations of \cvn\ used for Magnetic Doppler Imaging. 
\label{tbl:obs}}
\begin{tabular}{lcccrrrcccc}
\hline\hline
UT Date & \multicolumn{3}{c}{Stokes}     & \multicolumn{3}{c}{JD (2\,450\,000+)} & & & $S/N$\\
        & \multicolumn{3}{c}{Parameters} & $Q$~~~~~ & $U$~~~~~ & $V$~~~~~ & $\overline{\varphi}$ &  $\delta\varphi$ & range\\
%\noalign{\smallskip}
\hline
%\noalign{\smallskip}
 18 Feb 1997   &  $Q$ & $U$ & $V$ &  498.511 &  498.531 &  498.489 & 0.038 & 0.004 & 420--550 \\
 19 Feb 1997   &  $Q$ & $U$ & $V$ &  499.532 &  499.555 &  499.511 & 0.225 & 0.004 & 650--690 \\
 20 Feb 1997$^{\rm a}$   &      &     & $V$ &          &          &  500.422 & 0.387 &       & 550      \\
 21 Feb 1997   &  $Q$ & $U$ & $V$ &  501.490 &  501.504 &  501.470 & 0.582 & 0.003 & 270--320 \\
 22 Feb 1997   &  $Q$ &     & $V$ &  502.549 &          &  502.519 & 0.773 & 0.003 & 320--490 \\
 23 Feb 1997   &  $Q$ & $U$ & $V$ &  503.497 &  503.516 &  503.474 & 0.949 & 0.004 & 240--280 \\
 25 Feb 1997$^{\rm b}$ &      &     & $V$ &          &          &  505.581 & 0.331 &       & 470      \\
 25 Feb 1997   &  $Q$ & $U$ &     &  505.695 &  505.715 &          & 0.353 & 0.002 & 410--410 \\
 07 Feb 1998   &  $Q$ & $U$ & $V$ &  852.562 &  852.584 &  852.539 & 0.771 & 0.004 & 740--890 \\
 09 Feb 1998   &  $Q$ & $U$ & $V$ &  854.561 &  854.584 &  854.539 & 0.137 & 0.004 & 820--1020 \\
 11 Feb 1998   &  $Q$ & $U$ & $V$ &  856.545 &  856.567 &  856.524 & 0.499 & 0.004 & 910--1020 \\
 12 Feb 1998   &  $Q$ & $U$ & $V$ &  857.675 &  857.694 &  857.657 & 0.706 & 0.003 & 820--880 \\
 13 Feb 1998   &  $Q$ & $U$ & $V$ &  858.555 &  858.574 &  858.537 & 0.867 & 0.003 & 710--810 \\
 15 Feb 1998   &  $Q$ & $U$ & $V$ &  860.605 &  860.626 &  860.586 & 0.242 & 0.004 & 500--720 \\
 16 Feb 1998   &  $Q$ & $U$ & $V$ &  861.606 &  861.627 &  861.584 & 0.424 & 0.004 & 560--750 \\
 17 Feb 1998   &  $Q$ & $U$ & $V$ &  862.533 &  862.554 &  862.515 & 0.594 & 0.004 & 560--560 \\
 19 Feb 1998   &  $Q$ & $U$ & $V$ &  864.603 &  864.638 &  864.576 & 0.973 & 0.006 & 430--590 \\
 15 Jan 1999   &  $Q$ &     & $V$ & 1194.621 &          & 1194.600 & 0.310 & 0.002 & 400--450 \\
 19 Jan 1999   &  $Q$ & $U$ & $V$ & 1198.652 & 1198.672 & 1198.633 & 0.049 & 0.004 & 770--840 \\
 26 Jul 2004   &  $Q$ & $U$ & $V$ & 3213.362 & 3213.381 & 3213.400 & 0.413 & 0.004 & 440--480 \\
\noalign{\smallskip}
\hline
\end{tabular}
%\end{center}
%\begin{list}{}{}
%\item[]
\flushleft
Columns give the UT date for the beginning of the night when observation was obtained, the Stokes parameters observed (in addition to Stokes $I$), the Julian Dates of mid-exposure
for each observed Stokes parameter, the mean phase $\overline{\varphi}$ for all observed Stokes 
parameters, the maximum difference $\delta\varphi$ between $\overline{\varphi}$  and the phases of 
individual Stokes parameter observations, the range of the signal-to-noise
ratio per 2.6~km/s spectral pixel for two or three exposures. \\
$^{\rm a}$ The Stokes $Q$ and $U$ spectra are disregarded. See text for detailed explanation.\\
$^{\rm b}$ The Stokes $V$ observation 
%obtained on 25 Feb 1997 was 
is considered individually due to a larger-than-average phase difference with respect to the linear polarization spectra obtained on that night.
%\end{list}
\end{table*}

\section{Physical properties of \cvn}
\label{props}

Our modeling of the Stokes $IQUV$ spectra of \cvn\ and the interpretation of their rotational variability in terms of the surface maps of both magnetic field and chemical elements requires determining certain basic physical and geometrical properties of the target star:
\begin{enumerate}
\item[i]Atmospheric parameters -- \teff, \lgg\ and average chemical composition -- are needed as inputs for a model atmosphere code, which provides us with the vertical temperature-density stratification used for the numerical calculation of the local Stokes profiles and continuum intensities;
\item[ii]Rotation period $P_{\rm rot}$ and projected rotational velocity \vs\ are needed to establish the mutual phasing of observations and to evaluate the rotational Doppler shift across the stellar surface;
\item[iii]Two angles, $i$ and $\Theta$, establish the orientation of the stellar axis of rotation with respect to the observer \citep{piskunov:2002a}.
\end{enumerate}
Some of these parameters can be adjusted by optimizing the fit to the observations with multiple inversions \citep{kochukhov:2002c}. Other parameters produce little impact on the Stokes spectra and Doppler images and hence require a prior estimation.

\subsection{Atmospheric parameters}

\cvn\ (12 CVn A, HD\,112413, HR\,4915, HIP\,63125) was the first star given the ``A-type peculiar'' classification \citep{maury:1897} and is a prototype of spectrum variable stars. It was classified as A0pSiEuHg by \citet{cowley:1969} and was targeted by many notable magnetic and spectrum variability studies \citep{babcock:1952,pyper:1969,cohen:1970,borra:1977}, including early attempts to reconstruct maps of abundance spots and magnetic field with the Doppler Imaging technique \citep{goncharskii:1983,khokhlova:1984,glagolevskii:1985a}.

In the previous DI study of \cvn\ \citep{kochukhov:2002b} we have determined the effective temperature and surface gravity using theoretical model atmosphere fit to the optical spectrophotometry and hydrogen Balmer line profiles. We found that the {\sc ATLAS9} models computed with 10 times solar metal abundance provide a satisfactory fit to observations for \teff\,=\,$11600\pm150$~K and \lgg\,=\,$3.9\pm0.1$. Subsequently \citet{lipski:2008} used the same model atmosphere code to fit the optical spectral energy distribution simultaneously with the UV fluxes recorded by the IUE satellite. They showed that the UV data require \teff\,$\approx$\,10750~K, which is about 1000~K cooler than  \teff\ inferred from the optical spectrophotometry. The final \teff\,=\,$11250\pm500$~K recommended by \citet{lipski:2008} is formally not different from the temperature adopted by \citet{kochukhov:2002b}, but has a much larger uncertainty reflecting the difficulty of fitting the stellar spectrum in a broad wavelength region.

It is evident that further progress in deriving a realistic model structure of the atmosphere of \cvn\ can be achieved by computing models with individualised chemical composition and magnetic field \citep{shulyak:2004,kochukhov:2005a}. Nevertheless, for the purpose of our investigation, and to retain the possibility of a direct comparison with the results of the previous DI study, it is sufficient to use the model atmosphere parameters of \citet{kochukhov:2002b}, which represent reasonably well the mean atmospheric properties of the star. Using forward spectrum synthesis calculations we have verified that modification of \teff\ by $\pm500$~K produces only a marginal change of the line intensities, comparable or smaller than the uncertainty due to imperfect atomic data. Thus, the temperature uncertainty is not a critical factor in our analysis of \cvn.

In addition to the uncertainties mentioned above, the assumption of a constant model structure is not fully correct for spotted Ap stars since the line and continuous opacity is different inside and outside spots. However, this common assumption is justified for Doppler imaging studies similar to ours because the local line profiles are sensitive to model structure effects to a much smaller degree than to changes of abundance or magnetic field.

Taking into account the distance to \cvn, $d$\,=\,$35.2\pm1.1$ pc which follows from the revised Hipparcos parallax \citep{van-leeuwen:2007}, and adopting $V=2.89\pm0.04$ from \citet{pyper:1969}, we find the absolute visual magnitude $M_{\rm V}$\,=\,$0.16\pm0.08$. For the temperature \teff\,=\,$11600\pm500$~K and the bolometric correction $BC$\,=\,$-0.42$ measured by \citet{lipski:2008} we obtain $L$\,=\,$101\pm12$~$L_\odot$ and $R$\,=\,$2.49\pm0.26$~$R_\odot$ assuming $M^\odot_{\rm bol}$\,=\,4.75 for the Sun \citep{bessell:2000} and using an uncertainty of 0.1 for the bolometric correction of \cvn. Then, employing this stellar luminosity and temperature we can determine the evolutionary state of \cvn\ by comparing its observed position in the H-R diagram with the predictions of the theoretical evolutionary tracks by \citet{schaller:1992} and \citet{schaerer:1993}. Assuming an overall metallicity $Z=0.018$, we derive a stellar mass $M$\,=\,$2.97\pm0.07$~$M_\odot$ and an age of $1.65^{+0.6}_{-0.7}\times10^8$ years. These parameters yield the surface gravity \lgg\,=\,$4.12\pm0.09$, which can be brought in agreement with the spectroscopic estimate if the latter is corrected for the strong He deficiency of the atmosphere of \cvn\ \citep{kochukhov:2002b}.

The fundamental parameters of \cvn\ determined in our study, as well as those adopted from previous investigations, are summarized in Table~\ref{tbl:params}.

\begin{table}[!t]
%\begin{center}
\caption{Fundamental parameters of \cvn.
\label{tbl:params}}
%\centering
\begin{tabular}{lcc}
\hline\hline
Parameter & Value & Ref. \\
\hline
\teff & $11600\pm500$ K & (1), (2) \\
\lgg & $3.9\pm0.1$ & (1) \\
Distance & $35.2\pm1.1$ pc & (3) \\
$M_{\rm V}$ & $0.16\pm0.08$ & (4) \\
$BC$ & $-0.42\pm0.1$ & (2) \\
$L$ & $101\pm12$~$L_\odot$ & (4) \\
$R$ & $2.49\pm0.26$~$R_\odot$ & (4) \\
$M$ & $2.97\pm0.07$~$M_\odot$ & (4) \\
Age & $1.65^{+0.6}_{-0.7}\times10^8$ yr & (4) \\
\vs & $18.4\pm0.5$ \kms & (4) \\
$P_{\rm rot}$ & 5\fd46939 & (5) \\
$i$ & $120\pm5$\degr & (4) \\
$\Theta$ & $115\pm5$\degr & (4) \\
\hline
\end{tabular}
%\begin{list}{}{}
%\item[] 
\flushleft
References. (1) \citet{kochukhov:2002b}; (2) \citet{lipski:2008}; (3) \citet{van-leeuwen:2007}; (4) this work; (5) \citet{farnsworth:1932}.
%\end{list}
%\end{center}
\end{table}

\subsection{Rotational parameters}

The widely-used rotation period of \cvn, 5\fd46939, was determined by \citet{farnsworth:1932} based on a series of intensity measurements of the blue \ion{Eu}{ii} lines taken at several observatories during the years 1912--1932. Using this rotational period one can successfully phase together longitudinal magnetic field measurements obtained over the time span of 25 years \citep{wade:2000}. Thus, the rotational period of \citet{farnsworth:1932} is adequate for our analysis. We do not attempt to improve it since no recent extensive photometric or spectroscopic time-series observations are available for \cvn.

The projected rotational velocity is one of the key input parameters for DI inversions and needs to be determined with a relatively high precision \citep{kochukhov:2002c}. It is best found by optimizing the fit to line profiles observed at high spectral resolution. As described in Sect.~\ref{opt}, the intensity profiles of the strong \ion{Fe}{ii} and \ion{Cr}{ii} lines in the spectrum of \cvn\ are reproduced with \vs\,=\,$18.4\pm0.5$~\kms. This value is somewhat larger than $17.4\pm0.5$~\kms\ found by \citet{kochukhov:2002b} using \ion{Cr}{ii} lines of different strength. Although formally not significant, this discrepancy could reflect the difference in instrumental profile or, possibly, be an artifact of vertical chemical stratification neglected in our spectrum synthesis.

The orientation of the stellar rotational axis is characterized by the two angles $i$ and $\Theta$. The inclination, $0\degr\leq i \leq 180\degr$, is the angle between the rotational axis and the observer's line-of-sight. The values $i>90\degr$ correspond to the situation when we see a clockwise rotation of the star from the visible rotation pole. The azimuth angle, $0\degr\leq\Theta\leq 360\degr$ determines the sky-projected position angle of the rotational axis. This angle is counted counterclockwise from the North Celestial Pole \citep{landolfi:1993}. The azimuth angle is relevant and the distinction between $i$ and $180\degr-i$ can be made only when interpreting linear polarization observations. Note that since the Stokes $Q$ and $U$ parameters depend on trigonometric functions of $2\Theta$, the values of $\Theta$ and $\Theta+180\degr$ cannot be discriminated. 

\begin{figure}[!t]
\figps{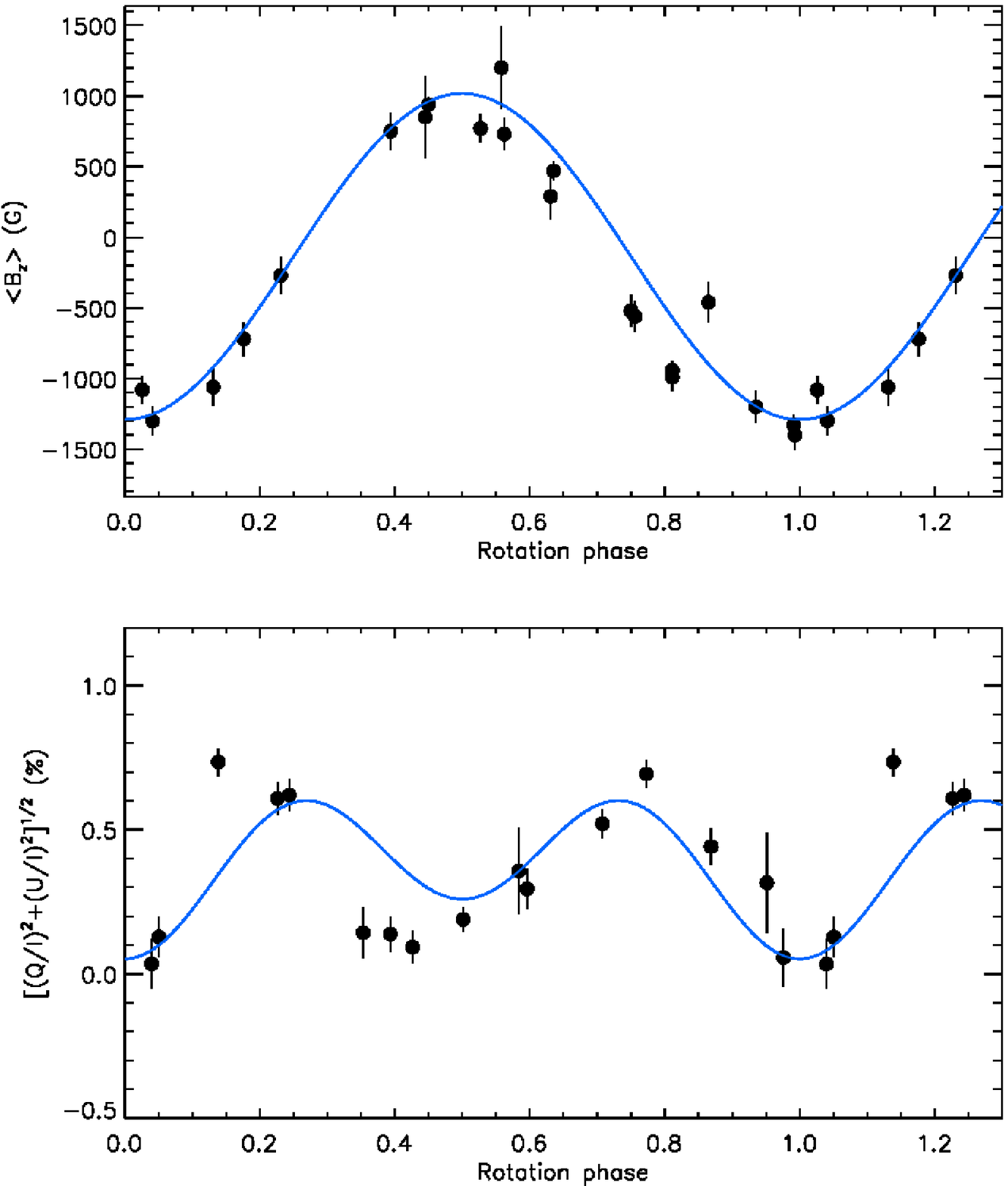}
\caption{Photopolarimetric measurements of the longitudinal magnetic field in \cvn\
(\citealt{borra:1977,landstreet:1982}, upper panel) and the total net
linear polarization derived from the LSD Stokes $Q$ and $U$ measurements of \citet{wade:2000}
(lower panel) compared with the predictions
of the dipolar model (solid curves) with parameters
$B_{\rm p}=4.6$~kG, $i=119$\degr\ and $\beta=78$\degr. The model with the same
$B_{\rm p}$ and $i^\prime=180\degr - i$, $\beta^\prime=180\degr - \beta$ yields
indistinguishable curves.}
\label{fig:dip}
\end{figure}

\begin{figure}[!t]
\figps{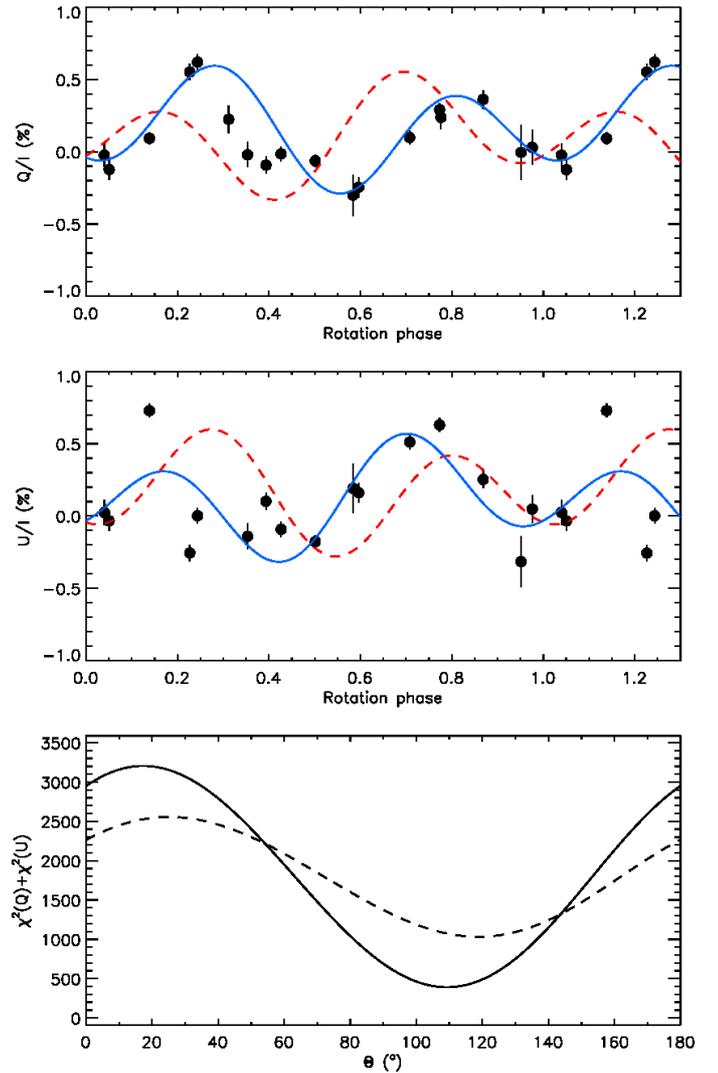}
\caption{Comparison of the net linear polarization (symbols) obtained from the LSD Stokes $Q$ and $U$ profiles
of \cvn\ (upper and middle panels, respectively) with the dipolar model predictions (lines).
The solid line corresponds to the magnetic field geometry given by $B_{\rm p}=4.6$~kG, $i=119\degr$,
$\beta=78$\degr, $\Theta=110$\degr. The dashed line shows predictions of the best-fitting model for the same
$B_{\rm p}$, $180\degr - i$, $180\degr - \beta$ and $\Theta=119$\degr.
The lower panel shows the total chi-square of the fit to net linear polarization curves as a function
of the azimuth angle $\Theta$ for the dipolar model with $i=119$\degr,
$\beta=78$\degr\ (solid line) and for the model with complementary angles
$180\degr - i$ and $180\degr - \beta$ (dashed line).
}
\label{fig:pqu}
\end{figure}

Numerical tests of the four Stokes parameter inversion procedure \citep{kochukhov:2002c} indicate that the tilt and, especially, the azimuth angle of the rotational axis can be reliably determined by minimizing the $\chi^2$ of the fit to observations in inversions with different $i$ and $\Theta$. However, our magnetic DI technique is fairly costly in computing time, which makes exploring all possible rotational axis orientations impractical. Instead we start by finding a plausible range for each angle using a simple dipolar field model and then refine the angles with a series of MDI reconstructions.

The inclination angle is inferred from the simultaneous fit of the dipolar field model to the Balmer line photopolarimetric longitudinal field measurements \citep{borra:1977,landstreet:1982} and the total linear polarization, $P_{\rm L}\equiv\sqrt{(Q/I)^2+(U/I)^2}$, obtained from the net linear polarization measurements of \citet{wade:2000}. We employ longitudinal field measurements obtained from H lines to avoid the strong modulation due to to abundance nonuniformities that is evident in longitudinal field measurements derived from, e.g., lines of Fe or Cr \citep{wade:2000}. The dipole model parameters include the polar field strength $B_{\rm p}$, magnetic obliquity $\beta$ and angles $i$ and $\Theta$. The theoretical $P_{\rm L}$ is computed following \citet{landolfi:1993}. Using the total linear polarization instead of individual net $Q$ and $U$ measurements allows us to avoid solving for $\Theta$ at this stage of the analysis. 

The comparison of the magnetic observables with the dipolar model predictions is presented in Fig.~\ref{fig:dip}. The scatter around the best-fit curve is considerable but the model evidently reproduces the main features of the \bz\ and $P_{\rm L}$ phase dependence if we adopt parameters $B_{\rm p}$\,=\,$4.6\pm0.5$~kG, $\beta$\,=\,$78\pm7\degr$, $i$\,=\,$119\pm12\degr$ (or $180\degr - \beta$\,=\,102\degr\ and $180\degr - i$\,=\,61\degr). The inclination angle obtained in this way is consistent with $i$\,=\,$53\pm8$\degr\ that can be derived from $P_{\rm rot}$, $R$ and \vs\ of \cvn\ using the usual oblique rotator relation.

In the second step we find the values of $\Theta$ which provide the best match to the observed rotational variation of the $Q$ and $U$ net linear polarization for the previously determined dipolar model parameters. Fig.~\ref{fig:pqu} demonstrates that a global $\chi^2$ minimum is found for $\Theta\approx110\degr$ and that $\beta=78\degr$, $i=119\degr$ provide a better match to the observations than
$180\degr - \beta$ and $180\degr - i$. The same conclusion was reached by \citet{wade:2000}. Thus, preliminary modeling suggests $i\approx120\degr$ and $\Theta\approx110\degr$. The final values of these parameters are established in Sect.~\ref{opt} with the help of multiple MDI inversions which take into account chemical inhomogeneities and incorporate deviations of the magnetic field topology from a simple dipole.

\section{Magnetic Doppler imaging}
\label{mdi}

\subsection{Magnetic inversion technique}
\label{method}

A detailed discussion of Magnetic Doppler Imaging using high-resolution circular and linear polarization spectra is provided by \citet{piskunov:2002a}. In that publication the MDI code \inv\ was introduced, and key numerical techniques employed in magnetic inversions were discussed. Here we briefly recount the basic principles of our MDI methodology, referring the reader to \citet{piskunov:2002a} for further details and for an in-depth description of technical issues.

The aim of the magnetic inversion is to derive the surface distributions of magnetic field and chemical abundances by minimizing the total discrepancy function
\beq
\Psi = \mathcal{D} + \mathcal{R},
\eeq
where $\mathcal{D}$ characterizes the discrepancy between the observed and computed phase-resolved spectra and $\mathcal{R}$ is the regularization functional. For DI with polarization data
\beq
\mathcal{D} = \sum_{k\varphi\lambda} \omega_k \left[ \mathcal{F}^{\rm comp}_{k\varphi\lambda}(\bb,\varepsilon^1,\varepsilon^2,...) - \mathcal{F}^{\rm obs}_{k\varphi\lambda}\right]^2/\sigma^2_{k\varphi\lambda},
\eeq
where $\mathcal{F}_{k\varphi\lambda}$ are the computed and observed Stokes parameter profiles for rotation phase $\varphi$ and wavelength point $\lambda$. The index $k$ represents summation over the available Stokes parameters (all four in the present study of \cvn), while weights $\omega_k$ are introduced to ensure approximately the same contribution of each Stokes parameter to $\mathcal{D}$. In practice this is achieved by choosing $\omega_k$ proportional to the phase-averaged amplitudes of the observed Stokes parameters. For our \cvn\ data set $\omega_I:\omega_Q:\omega_U:\omega_V=1:45:45:6$.

Theoretical Stokes $IQUV$ spectra for a given rotation phase depend on the surface topology of the magnetic field $\bb$ and the geometry of the abundance distributions $\varepsilon^1$, $\varepsilon^2$, etc. Normalized Stokes profiles are obtained by summing the Doppler-shifted local Stokes spectra $\mathcal{X}^{(k)}_i$ over a discrete surface grid and dividing by the disk-integrated intensity in the unpolarized continuum
\beq
\mathcal{F}^{\rm comp}_{k\varphi\lambda}=\sum_{i=1}^{N_\varphi} S^{(\varphi)}_i \mathcal{X}^{(k)}_i(\bb,\varepsilon^1,\varepsilon^2,\dots,\Delta\lambda_{\rm D})/\sum_{i=1}^{N_\varphi} S^{(\varphi)}_i I^c_i.
\eeq
Here, the index $i$ runs over a set of $N_{\varphi}$ surface zones visible to the observer at the rotational phase $\varphi$ and $S^{(\varphi)}_i$ is the projected area of surface element $i$. For the present study of \cvn\ we use a grid with 695 surface zones, which is fully sufficient given the moderate \vs\ of the star and the limited spectral resolution of the observational data. The local Stokes parameter profiles $\mathcal{X}^{(k)}_i$ are computed for a given model atmosphere and local values of the abundances and magnetic field by solving numerically the polarized radiative transfer equation in its general form, as described by \citet{piskunov:2002a}. Thus, theoretical spectra are computed by \inv\ fully self-consistently with the current magnetic and abundance maps and treating the line formation in the magnetized stellar atmosphere without any of the simplifying assumptions that are commonly used, e.g. a Milne-Eddington atmosphere or fixed local Gaussian line profiles.

We apply regularization to ensure stability of the complex optimization process of magnetic DI and to obtain a unique solution independent of the initial guess and of the surface discretization. The general form of the regularization functional $\mathcal{R}$ implemented in \inv\ is 
\begin{eqnarray}
\label{eq:reg}
\mathcal{R} & = & \Lambda_1 \sum_{i} \sum_j (\bb_i - \bb_j)^2 + \Lambda_2 \sum_{i} (\bb_i - \bb^{\rm mult}_i)^2 \\
&+& \Lambda_3 \left[\sum_{i} \sum_j (\varepsilon^1_i - \varepsilon^1_j)^2
                   +\sum_{i} \sum_j (\varepsilon^2_i - \varepsilon^2_j)^2 + \dots\right]. \nonumber
\end{eqnarray}
The first and third terms represent the Tikhonov regularization functions for magnetic field and abundance distributions, respectively. $\Lambda_1$ and $\Lambda_3$ are corresponding regularization parameters, selected in such a way that $\mathcal{D}/\mathcal{R}$\,$\approx$\,2--10 at the convergence. Application of the Tikhonov regularization leads to local smoothing of the solution by minimizing the difference between the value of parameters in the grid point $i$ and all neighbouring points $j$. Appropriate choice of the Tikhonov regularization ensures that the magnetic and abundance distributions reconstructed by our code represent the simplest possible solution which is still compatible with observational data. At the same time, the Tikhonov regularization does not impose an \textit{a priori} mathematical model for the global magnetic field geometry or abundance distribution.

A comprehensive series of numerical experiments with \inv\ was presented by \citet{kochukhov:2002c}. This paper explored the performance of the magnetic DI method for different combinations of stellar parameters, magnetic field strengths and geometries by reconstructing surface maps from simulated Stokes parameter observations. These numerical tests convincingly demonstrate that a successful inversion, revealing magnetic field structures at both the large and small scales, is possible when the code is applied to the full Stokes vector data set in conjunction with Tikhonov regularization. In fact, \citet{piskunov:2005} showed that, from the basic mathematical standpoint, the underlying inverse problem of magnetic mapping with all four Stokes parameters is well-posed and has a unique solution. In this case regularization is only needed to suppress instabilities produced by the noise in real data and its sparse phase and wavelength coverage. However, due to the lack of high-quality Stokes $IQUV$ stellar spectra, the only previous magnetic DI inversion based on polarization observations in all four Stokes parameters was our study of the Ap star \cam\ \citep{kochukhov:2004d}.

While Stokes $IQUV$ datasets suitable for Magnetic DI are relatively rare, comparable Stokes $IV$ datasets are more common. Magnetic DI with only circular polarization observations has therefore been more widely applied. However, such datasets result in maps which are less informative and robust compared to those computed from full Stokes vector data. The $IV$ mapping problem is intrinsically ill-conditioned and the outcome of such inversions depends sensitively on the choice of regularization function and the initial guess. Our numerical experiments with \inv\ \citep{piskunov:2002a,kochukhov:2002c} showed that a successful reconstruction of the global field topology of Ap stars using only Stokes $IV$ data requires imposing a prior assumption that strongly restricts the range of possible derived magnetic field topologies. In our DI code this assumption is implemented in the form of so-called multipolar regularization (the second term in Eq.~(\ref{eq:reg})), which directs the solution toward a general, non-axisymmetric second-order multipolar expansion of the magnetic field structure. The multipolar model field $\bb^{\rm mult}$, calculated by our code at each iteration, is equivalent to the dipole plus quadrupole parameterization of \citet{bagnulo:1996}. Stokes $IV$ time series of the Ap stars \cvn\ \citep{kochukhov:2002b}, HD\,24712 \citep{luftinger:2010} and HD\,72106 \citep{folsom:2008} were interpreted using \inv\ in the multipolar regularization mode. All these studies inferred basically dipolar field geometries for the target stars.

The availability of \mus\ four Stokes parameter time-resolved spectra for \cvn\ provides us with the capability to perform magnetic inversion constrained only by local, Tikhonov regularization ($\Lambda_1\ne0$, $\Lambda_2=0$, see Sect.~\ref{stok4}). At the same time, it is also very instructive to compare these ultimate MDI results with the maps obtained disregarding linear polarization observations and globally constraining the magnetic inversion with the multipolar regularization. In Sect.~\ref{stok2} we present results for magnetic DI in the circular polarization mode, with $\Lambda_1=0$, $\Lambda_2\ne0$.

\begin{table}[!t]
%\begin{center}
\caption{Atomic parameters of spectral lines employed in magnetic inversions.
\label{tbl:lines}}
%\centering
\begin{tabular}{llrlcccc}
\hline\hline
Ion &~~~$\lambda$ (\AA) & $E_{\rm lo}$ (ev) &~~$\log gf$ & $J_{\rm lo}$ & $J_{\rm up}$ & $g_{\rm lo}$ & $g_{\rm up}$ \\
\hline
\ion{Cr}{ii} & 4824.127$^{\rm a}$ &  3.871 & $-0.970$ & 4.5 & 4.5 & 1.34 & 1.34 \\
\ion{Fe}{ii} & 4824.198           & 10.288 & $-1.215$ & 4.5 & 4.5 & 1.45 & 1.22 \\
\ion{Fe}{ii} & 4824.838           &  8.145 & $-1.893$ & 4.5 & 3.5 & 1.25 & 1.24 \\
\ion{Fe}{ii} & 4923.927           &  2.891 & $-1.320$ & 2.5 & 1.5 & 2.00 & 2.40 \\
\ion{Fe}{ii} & 5018.440$^{\rm a}$ &  2.891 & $-1.271$$^{\rm b}$ & 2.5 & 2.5 & 2.00 & 1.87 \\
\ion{Fe}{ii} & 5019.462           &  5.569 & $-2.697$ & 3.5 & 4.5 & 1.14 & 1.10 \\
\hline
\end{tabular}
\flushleft
The columns specify the ion, the central wavelength $\lambda$,
excitation potential of the lower atomic level $E_{\rm lo}$, quantum numbers $J$ and Land\'e factors
$g$ for both the lower and upper levels.\\
$^{\rm a}$ Dominant blend components.\\
$^{\rm b}$ Oscillator strength is adjusted in the MDI inversion.\\
\end{table}

\subsection{The choice of spectral lines}
\label{lines}

The amplitude of linear polarization signatures in metal lines of even the most strongly-magnetic Ap stars is typically at the level of only $10^{-3}$ \citep{wade:2000b}. Consequently, we are limited to studying the Stokes $Q$ and $U$ profiles of a few deepest, magnetically sensitive individual lines in the stellar spectra. Using such features is usually not optimal for abundance mapping because the lines are saturated over a significant part of the stellar surface and thus their variation is subdued in comparison to weaker lines of the same ions. For this reason, and also due to the relatively low resolving power of the \mus\ observations, we do not expect to achieve the same quality of abundance inversions as demonstrated by \citet{kochukhov:2002b} in the previous DI analysis of \cvn. However, this reduced sensitivity of the strong lines to abundance inhomogeneities make them more suitable for our primary goal of the mapping magnetic field using four Stokes parameter profiles.

We found three lines in the spectrum of \cvn\ appropriate for magnetic DI. The two \ion{Fe}{ii} features at $\lambda$~4923.93 and 5018.44~\AA\ were previously used by \citet{kochukhov:2004d} and \citet{khalack:2006} in the studies of other Ap stars observed with the \mus\ spectropolarimeter and proved to be effective for the reconstruction of the magnetic field topology. Their mean Land\'e factors, $\bar{g}=1.7$--1.9, are among the largest for metal lines of such intensity. Both lines show a clear and complex signal in the Stokes $Q$ and $U$ spectra, detected at high significance levels for the majority of rotation phases (see Fig.~\ref{fig:prf_smooth}). We also use the \ion{Cr}{ii} 4824.13~\AA\ line for magnetic mapping. This line shows less prominent Stokes $Q$ and $U$ profiles because it is somewhat weaker than the \ion{Fe}{ii} lines, and exhibits a smaller Zeeman effect ($\bar{g}=1.3$). Nevertheless, simultaneous recovery of the surface magnetic field distribution from the lines of more than one chemical element always improves the robustness of MDI \citep{kochukhov:2002c}.

The complete line list adopted in the magnetic inversions of \cvn\ is presented in Table~\ref{tbl:lines}. Parameters of the two strong \ion{Fe}{ii} lines, one \ion{Cr}{ii} line and three weaker \ion{Fe}{ii} blends are extracted from the VALD database \citep{kupka:1999}. The oscillator strength of the lines from \ion{Fe}{ii} multiplet 42, to which the 4923.93 and 5018.44~\AA\ lines belong, are notorious for the large scatter of $\log gf$ recommended in different literature sources. Even the relative $\log gf$ values of these lines are not known with the precision necessary for detailed spectrum synthesis modeling, prompting \citet{kochukhov:2004d} to perform separate MDI inversions for each of these lines in their study of \cam. To overcome this problem here we modified \inv, adding the oscillator strength of the \ion{Fe}{ii} 5018.44~\AA\ line to the list of free parameters determined by the code. Table~\ref{tbl:lines} lists the final $\log gf$ for this lines, which is 0.051~dex lower than the value recommended by VALD.

Additional line data required for calculation of the Zeeman splitting includes Land\'e factors of the upper and lower atomic levels and corresponding $J$ quantum numbers. These quantities are also specified in Table~\ref{tbl:lines}, with Land\'e factors originating from the theoretical computations by \citet{kurucz:1993a} and extracted using the VALD interface.

\begin{figure}[!t]
\figps{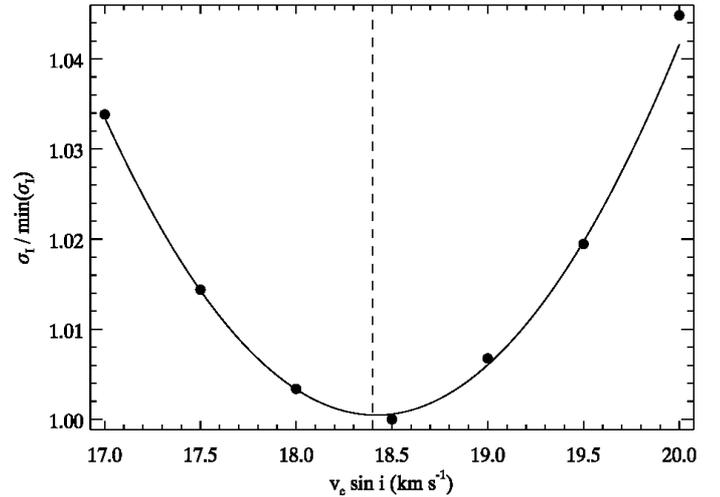}
\caption{The relative standard deviation of the fit to \ion{Cr}{ii} and \ion{Fe}{ii} Stokes $I$ profiles is plotted as a function of \vs\ (symbols). The solid curve shows the quadratic fit in the
17.5--19.5~\kms\ region. The vertical dashed line corresponds to the optimal projected rotational velocity \vs\,=\,$18.4\pm0.5$~\kms.}
\label{fig:vsini}
\end{figure}

\begin{figure}[!th]
\figps{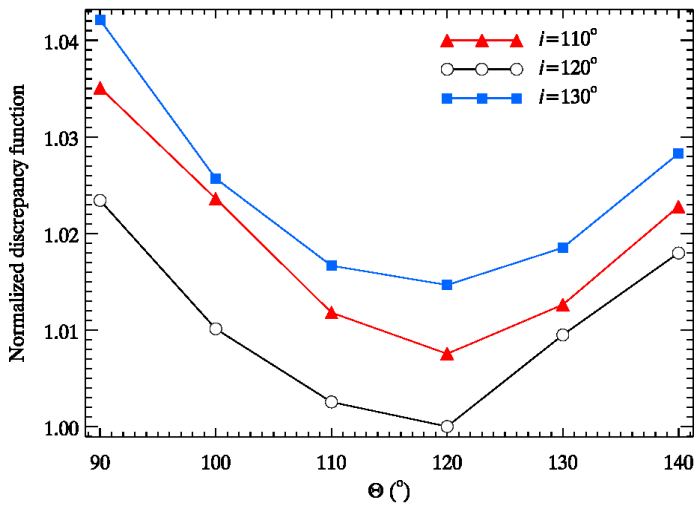}
\caption{The relative total discrepancy function 
%obtained in the Stokes $IQUV$ inversions
is plotted against the azimuth angle $\Theta$ 
adopted for the inversion.
%of the stellar rotation axis. 
Different symbols show results obtained for three values of the inclination
angle $i$. From these curves we infer the optimal values $\Theta=115\pm5$\degr\ and
$i=120\pm5$\degr.}
\label{fig:it}
\end{figure}

\subsection{Optimization of \vs, $i$ and $\Theta$}
% and orientation of the rotational axis}
\label{opt}

Magnetic inversion is sensitive to the adopted projected rotational velocity and the three-dimensional orientation of the stellar rotational axis. Numerical experiments presented by \citet{kochukhov:2002c} showed that solutions with incorrect values of these parameters tend to have a higher $\chi^2$ of the fit to the observed Stokes profiles. For \vs\ this degradation of the fit quality is most clearly seen in Stokes $I$. A measure of the final discrepancy between the observed and computed intensity profiles of the \ion{Cr}{ii} and \ion{Fe}{ii} lines is illustrated in Fig.~\ref{fig:vsini}. Examining the inversion results for 7 trial values of \vs\ in the range from 17 to 20~\kms\ we find a minimum for \vs\,=\,18--19~\kms. A parabolic fit to these results yields a minimum at \vs\,=\,$18.4\pm0.5$~\kms. We adopt this rotational velocity for all subsequent inversions presented in our paper.

We followed a similar strategy to optimize the choice of the inclination and azimuth angles. The modeling of the mean longitudinal field and net linear polarization suggests that these angles are close to 120\degr\ and 110\degr, respectively. To determine the values of $i$ and $\Theta$ we carried out inversions for the ranges $i=110$--130\degr\ and $\Theta=90$--140\degr\ with a 10\degr\ step for both angles. The normalized total discrepancy function obtained by \inv\ at the convergence is illustrated in Fig.~\ref{fig:it} for the entire grid of 18 inversions. It is evident that $i=120\degr$ gives a smaller discrepancy than either $i=110\degr$ or 130\degr, while the optimal $\Theta$ is constrained to the 110--120\degr\ interval independent of inclination. Finally, we establish $i=120\pm5\degr$ and $\Theta=115\pm5\degr$ and adopt this orientation of the stellar rotation axis for all magnetic inversions presented below.

\section{Results}
\label{results}

\subsection{Full Stokes vector inversion}
\label{stok4}

\begin{figure*}[!th]
\fifps{17.5cm}{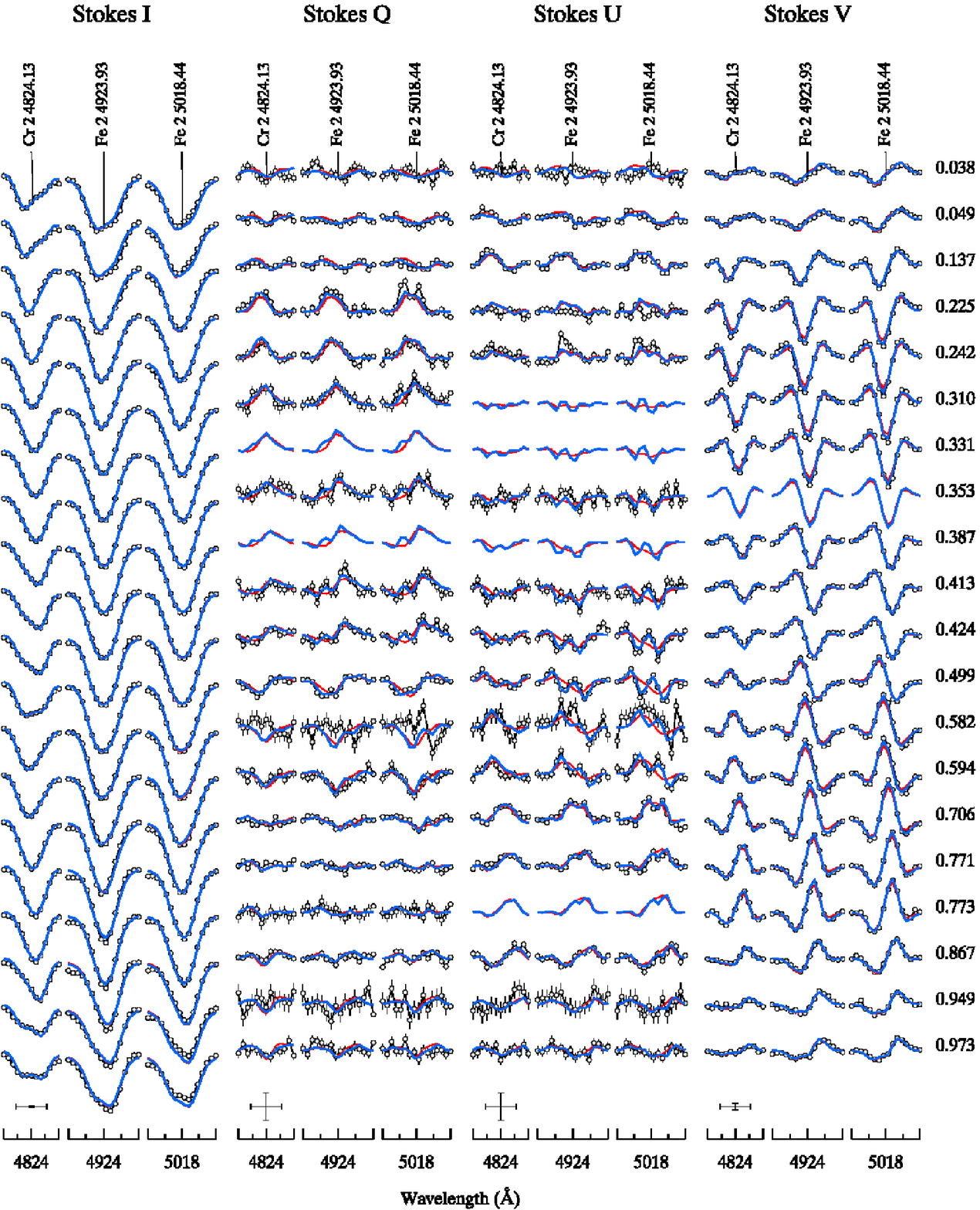}
\caption{The comparison between observed (symbols) and synthetic (solid curves) four
Stokes parameter spectra of \cvn. The thick curves show the best fit to the Fe and Cr lines obtained
with the four Stokes parameter DI modeling of the magnetic field geometry (Fig.~\ref{fig:fld_best})
and horizontal abundance distributions of both elements (Fig.~\ref{fig:abn}).
The thin lines show theoretical Stokes profiles illustrating the outcome of the MDI inversion with 
Tikhonov regularization enhanced by a factor of 10 relative to its optimal value (Fig.~\ref{fig:fld_smooth}). 
%
%corresponding to the best-fit magnetic geometry reconstructed in the Stokes
%$IV$ imaging with multipolar regularization (Fig.~\ref{fig:fld_mult}).
%
Spectra for consecutive rotational phases are shifted in
the vertical direction for display purposes.  Rotational phases are indicated in the column to the right of the
Stokes $V$ panel. The bars at the lower left of each panel show the horizontal and vertical
scale (0.5~\AA\ and 1\% of the Stokes $I$ continuum intensity respectively). The number in
brackets at the top of each panel quotes the mean deviation between observations and spectrum
synthesis for that Stokes parameter.}
\label{fig:prf_smooth}
\end{figure*}

\begin{figure*}[!th]
\figps{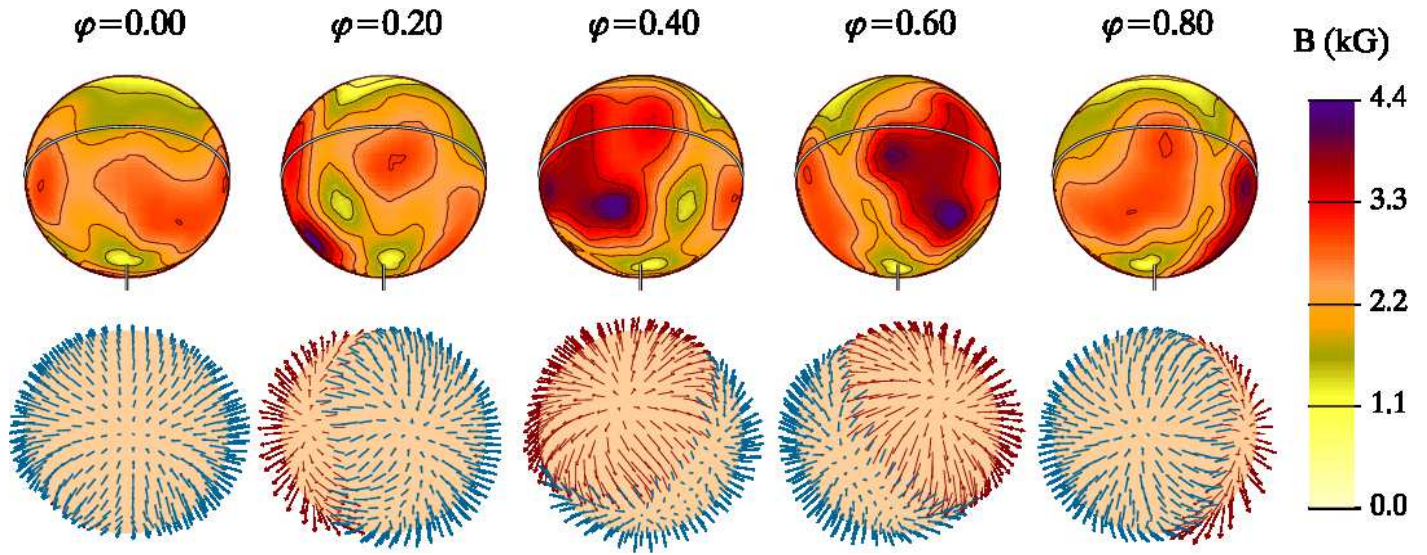}
\caption{Surface magnetic field distributions of \cvn\ derived from the Stokes $IQUV$
profiles of the \ion{Fe}{ii} and \ion{Cr}{ii} lines. The star is shown at five equidistant rotational
phases as indicated at the top of the figure. The aspect corresponds to the inclination angle $i=120\degr$ and vertically oriented rotational axis.
The upper row of spherical plots visualizes the distribution of field strength, with contours of equal magnetic field strength are plotted every 0.5~kG. 
The thick line shows the stellar rotational equator. Rotational axis is indicated with the vertical bar.
The lower panel shows the orientation of the magnetic vectors. In these vector maps the black arrows show field vectors pointing outside
the stellar surface and the light arrows correspond to the vectors pointing inwards. The
arrow length is proportional to the field strength.}
\label{fig:fld_best}
\end{figure*}

\begin{figure*}[!th]
\figps{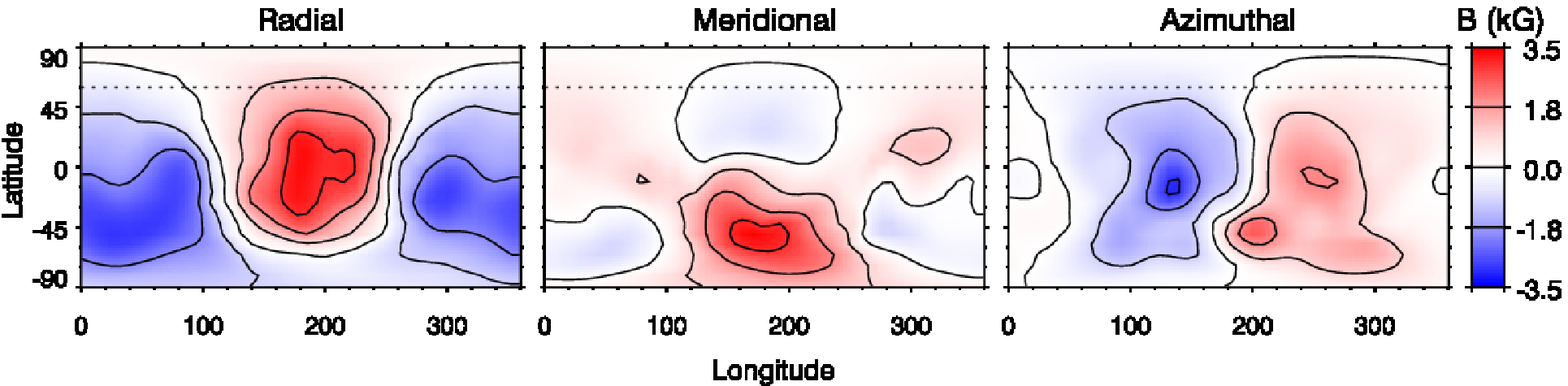}
\caption{Rectangular projection of the magnetic field geometry of \cvn\ derived from the Stokes $IQUV$ profiles of the \ion{Fe}{ii} and \ion{Cr}{ii} lines. The longitude coordinate divided by 360\degr\ corresponds to the phase of the subsolar meridian in Fig.~\ref{fig:prf_smooth}. The dotted line shows the highest visible latitude for the adopted inclination angle $i=120\degr$. The lines of equal field strength are shown for the [$-3$,$+3$]~kG interval, with a 1~kG step.}
\label{fig:fld_best_rect}
\end{figure*}

We reconstruct the magnetic field topology and surface distributions of Cr and Fe in a simultaneous inversion, using 20 phases of observations in all four Stokes parameters. The best-fitting theoretical profiles obtained by \inv\ at the convergence are shown in Fig.~\ref{fig:prf_smooth} with the thick solid line. Evidently, the observed line profile variability in the intensity and polarization is satisfactorily reproduced. The only minor systematic discrepancy is seen in the profile of the \ion{Fe}{ii} 5018.44~\AA\ line, which is overestimated in the theoretical spectra corresponding to the rotation phases around negative magnetic extremum ($\varphi$\,=\,0.949--0.049). The linear polarization profiles are well-fitted by \inv. Occasional discrepancies between the observations and the model spectra (e.g., Stokes $Q$ profile of \ion{Fe}{ii} 5018.44~\AA\ for $\varphi$\,=\,0.225) are always limited to only one out the three lines studied and thus is likely due to observational uncertainties. Especially impressive is the ability of the model to reproduce the systematic details of the linear polarization profiles, such as the peculiar "w"-shape of Stokes $U$ around phase 0.4. The agreement between the observations and synthetic Stokes $V$ spectra is excellent.

The surface magnetic field distribution of \cvn\ recovered by \inv\ from the phase-resolved Stokes profiles of the \ion{Cr}{ii} and \ion{Fe}{ii} lines is presented in Fig.~\ref{fig:fld_best}. This plot shows spherical projections of the magnetic map, displaying separately the field strength and the field orientation distributions. In this and similar plots the star is shown at 5 equidistant rotational phases, at the aspect angle corresponding to $i=120\degr$ and $\Theta=0\degr$. The overall structure of the magnetic field in \cvn\ is dipolar-like in the sense that approximately half of the stellar surface is covered with the outward-directed radial field while another half exhibits inward-directed field. This global magnetic field topology agrees with the orientation of the dipolar axis suggested in previous magnetic analyses of \cvn\ where multipolar models were fitted to integral magnetic observables \citep{borra:1977,gerth:1999} or to Stokes $IV$ line profiles \citep{kochukhov:2002b}. However, the field strength distribution illustrated in the upper panel of Fig.~\ref{fig:fld_best} reveals that the derived magnetic field topology of \cvn\ resembles that of a dipole on the largest spatial scale. The outcome of our four Stokes parameter inversion suggests that the field is, in fact, far more complex. In particular, there is a definite asymmetry in the field strength and structure between the negative magnetic pole ($\varphi\approx0.0$) and the positive one ($\varphi\approx0.5$). The field is clearly stronger in the latter case and its structure is dominated by the high-contrast magnetic spots where the field strength reaches 4.5~kG locally. This is approximately 1~kG higher than the local field at the opposite side of the star. 

The rectangular projection of the magnetic field distribution reconstructed for \cvn\ is presented in Fig.~\ref{fig:fld_best_rect} for all three components of the field vector. This figure again gives an impression of the global dipolar topology locally distorted by small-scale features. The radial field component is most similar to the surface magnetic structure expected for an oblique dipole, while the meridional and azimuthal components deviate significantly from a simple, low-order multipolar shape.

\begin{figure*}[!th]
\figps{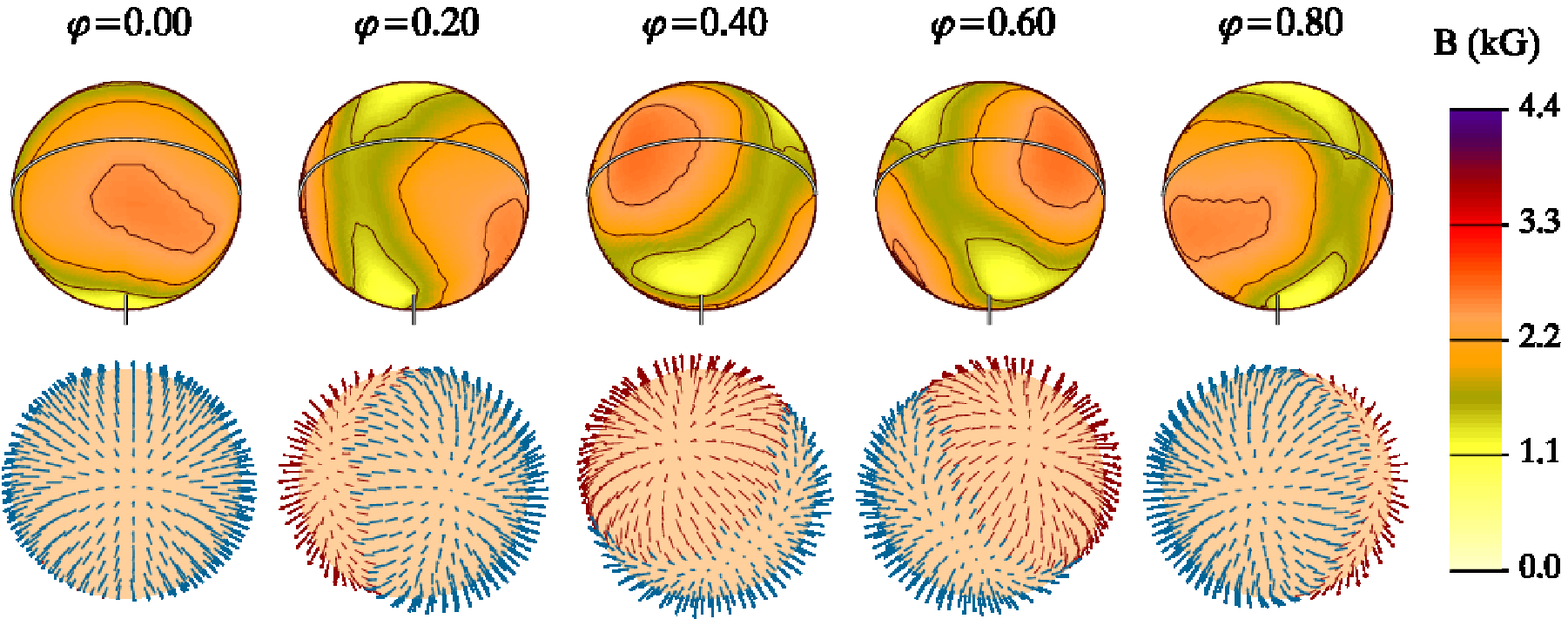}
\caption{Same as Fig.~\ref{fig:fld_best} but for the Stokes $IQUV$ imaging with ten times larger
Tikhonov regularization for the magnetic field.}
\label{fig:fld_smooth}
\end{figure*}

Is it possible that the complexity seen in the MDI maps of \cvn\ is an artifact produced by the noise in observations combined with an insufficiently strong regularization adopted in the magnetic mapping? We have thoroughly examined this possibility by conducting inversions with different values of the Tikhonov regularization parameter for magnetic field, $\Lambda_1$. These tests confirm that our choice of regularization is correct because any substantial increase of the smoothing of the magnetic maps by the Tikhonov regularization noticeably worsens the agreement between theoretical spectra and the linear polarization observations. For example, with the Tikhonov regularization increased by a factor of 10 the code converges to a markedly simpler field topology, with the field strength not exceeding 3~kG (Fig.~\ref{fig:fld_smooth}). The resulting fit quality is unchanged for Stokes $I$ and is only slightly worse for Stokes $V$ relative to the profiles corresponding to the complex field model. However, the smoothed field structure fails to reproduce the Stokes $Q$ and $U$ spectra at phases $\varphi$\,=\,0.353--0.594, when the most complex part of the surface magnetic field distribution crosses the line of sight.  For all these phases the observed linear polarization profiles exhibit a double-wave structure with two minima and two maxima across each spectral line (e.g. the "w"-shaped profiles of Stokes $U$). On the other hand, these features are well-reproduced by the best-fitting complex magnetic field distribution. In contrast, theoretical profiles corresponding to the smoothed magnetic map lack the double-wave structure. The resulting degradation of the fit quality for the $Q$ and $U$ spectra is readily seen in the increase of $\chi^2$ by 30--60\%.

This analysis shows that the small-scale features in our magnetic map of \cvn\ are directly connected to the particular morphology of the linear polarization signatures, observed consistently in all spectral lines studied. Thus, we can assert that our final magnetic field topology of \cvn\ (Fig.~\ref{fig:fld_best}) is truly the simplest possible field structure consistent with the available Stokes parameter observations. Therefore, the inferred complexity and substantial local deviation of the stellar magnetic geometry from the low-order multipolar field are real. It is remarkable that we are able to detect and characterize these complex magnetic structures using Stokes $IQUV$ data, but not using Stokes $IV$ data.

\begin{figure*}[!th]
\figps{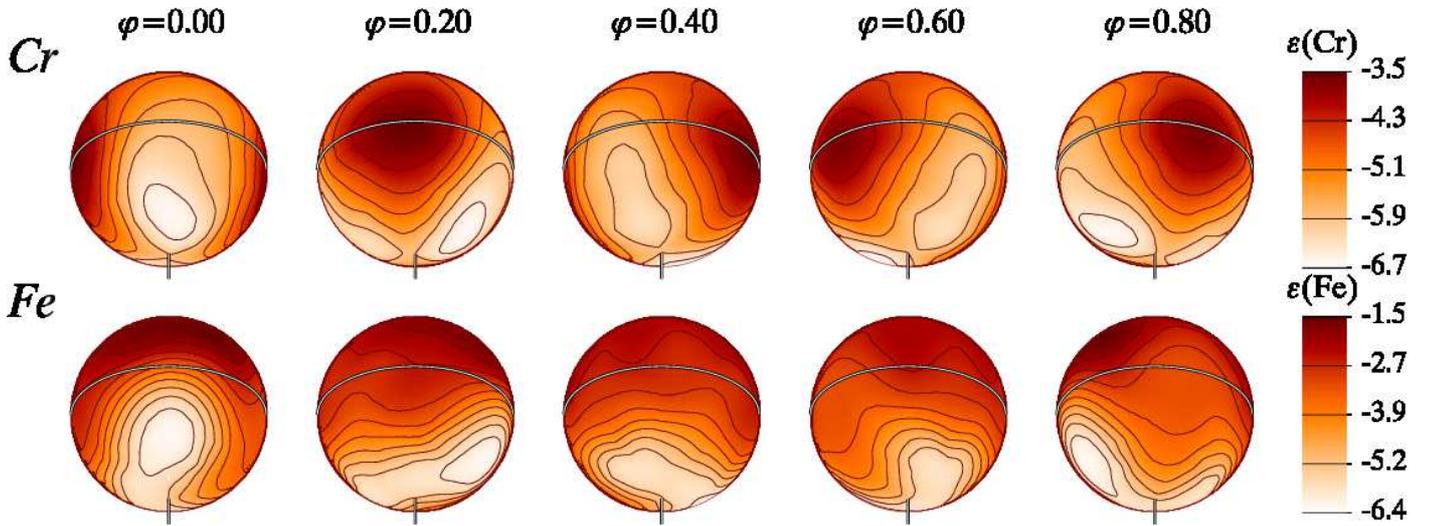}
\caption{Surface distribution of the Cr and Fe abundances derived simultaneously with the
magnetic field geometry in Fig.~\ref{fig:fld_best}.
The star is shown at five equidistant rotational
phases as indicated at the top of the figure. The aspect corresponds to the inclination angle
$i=120\degr$ and vertically oriented rotational axis. The element concentrations are given in the logarithmic units relative to the total atomic number density: $\varepsilon(\mathrm{El})\equiv\log{(N_{\rm El}/N_{\rm tot})}$.The contour lines in the spherical abundance
maps are plotted with a step of 0.5~dex. The thick line shows the stellar rotational equator. Rotational
axis is indicated with the vertical bar.}
\label{fig:abn}
\end{figure*}

The surface abundance maps of Cr and Fe reconstructed by \inv\ simultaneously with the magnetic field topology are presented in Fig.~\ref{fig:abn}. We infer a high-contrast distribution for both elements, with the abundance varying by 3--4~dex and pronounced abundance minima at phase $\varphi$\,=\,0. These abundance images are qualitatively similar to the iron-peak element maps reconstructed by \citet{kochukhov:2002b} but have a lower surface resolution for the reasons discussed in Sect.~\ref{lines}. In addition, the map obtained from the two strong \ion{Fe}{ii} lines exhibits noticeably smaller abundance values in the region around the negative magnetic pole ($\log{N_{\rm Fe}/N_{\rm tot}<-6}$ compared to $\log{N_{\rm Fe}/N_{\rm tot} \geq -4.5}$ obtained by \citealt{kochukhov:2002b}). This can possibly be ascribed to the effects of vertical chemical stratification, which we do not take into account in our analysis. A vertical abundance distribution with a transition from a high abundance in the deeper atmosphere to a lower element concentration in the higher layers is often observed in cooler Ap stars \citep[e.g.,][]{kochukhov:2006b,kochukhov:2009a}. If such stratification is present in some parts of the surface of \cvn, it would lead to substantial weakening of the cores of intrinsically strong \ion{Fe}{ii} lines relative to intrinsically weak lines, yielding a smaller abundance if such lines are analyzed neglecting chemical stratification. However, these effects generally do not significantly influence polarization profiles and thus a detailed treatment of vertical abundance inhomogeneities is beyond the scope of our study.

\subsection{Stokes $IV$ inversion with multipolar regularization}
\label{stok2}

%\onlfig{6}{
\begin{figure*}[!th]
\fifps{17.5cm}{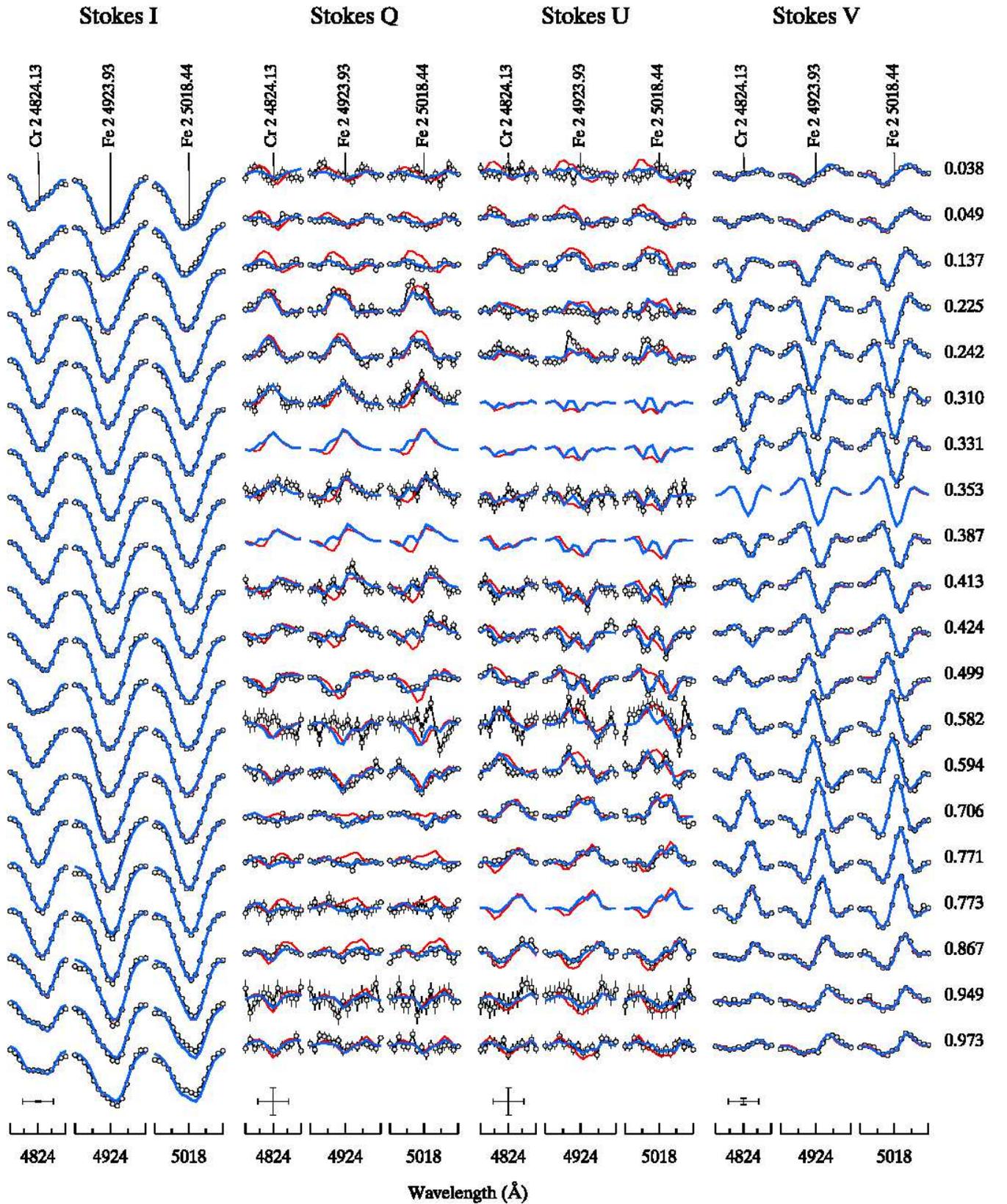}
\caption{Same as Fig.~\ref{fig:prf_smooth} except that theoretical four Stokes parameter spectra shown with
thin lines correspond to the best-fit magnetic geometry reconstructed in the Stokes
$IV$ imaging with multipolar regularization.
%illustrate the outcome of the MDI inversion with Tikhonov regularization enhanced by a factor
%of 10 relative to its optimal value. 
The corresponding magnetic field geometry of \cvn\ is presented in
Fig.~\ref{fig:fld_mult}.
}
\label{fig:prf_mult}
\end{figure*}
%}

\begin{figure*}[!th]
\figps{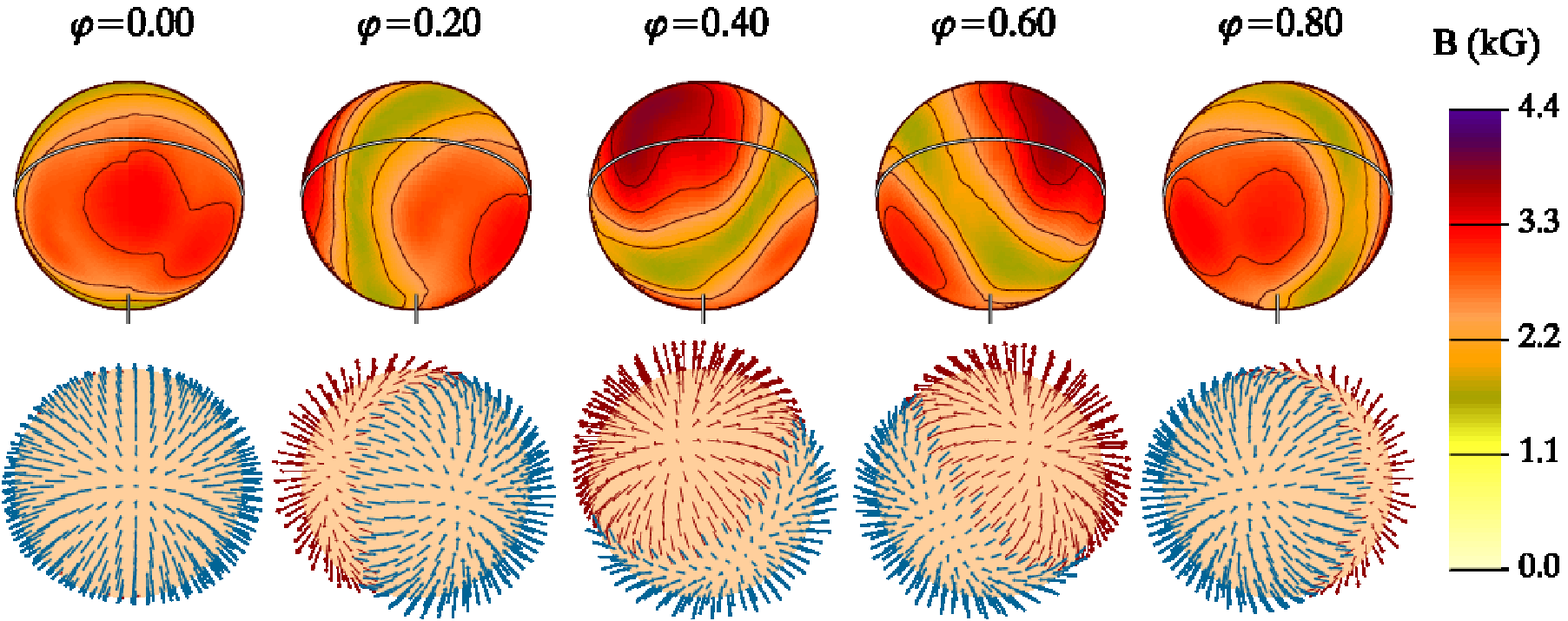}
\caption{Same as Fig.~\ref{fig:fld_best} but for the Stokes $IV$ imaging with multipolar
regularization.}
\label{fig:fld_mult}
\end{figure*}

Magnetic inversion in all four Stokes parameters, similar to our analysis of \cvn, remains very uncommon in the research field of stellar magnetism due to the considerable difficulties associated with the acquisition of the necessary observational data. Instead, magnetic DI and Zeeman Doppler Imaging (ZDI) studies employing time-resolved Stokes $IV$ spectra are much more widely applied to study field topologies of magnetic Ap stars and active late-type stars. In this section we assess how much of the complexity of the magnetic field of \cvn, evident from the full Stokes vector reconstruction, can be retrieved in the analysis of an incomplete, circular polarization-only data set. As described in Sect.~\ref{method}, in this experiment we use \inv\ in Stokes $IV$ inversion mode with the multipolar regularization. We carry out magnetic field reconstruction for the same three Cr and Fe spectral features, although in principle the number of usable spectral lines is much larger if we are not concerned with the analysis of the Stokes $Q$ and $U$ spectra.

The fit to the $I$ and $V$ spectra of \cvn\ and prediction for the linear polarization profiles obtained by our code at the convergence of the $IV$ mapping problem is illustrated with the thin line in Fig.~\ref{fig:prf_mult}. To enable a direct comparison with the previous results, this figure also shows the theoretical profiles for our optimal four Stokes parameter inversion (same as in Fig.~\ref{fig:prf_smooth}). At the resolution of the \mus\ spectra we see no difference in the intensity and circular polarization between the two sets of synthetic profiles. Yet the picture is very different for Stokes $Q$ and $U$. The model topology inferred from the circular polarization alone evidently yields systematically higher amplitude of the linear polarization profiles and cannot match the complex shape of the observed $Q$ and $U$ signal around rotational phase $\varphi\approx0.5$. This degradation of the fit quality is readily seen with a factor of 2--3 increase of the $\chi^2$ for both $Q$ and $U$ spectra. Thus, a successful description of the phase variation of the intensity and circular polarization spectra gives no guarantee that the resulting model of the stellar surface magnetic field is also adequate for the Stokes $Q$ and $U$ data.

The magnetic map derived in the Stokes $IV$ inversion is shown in Fig.~\ref{fig:fld_mult}. This field structure is very close to a dipolar topology with a mild non-axisymmetric quadrupolar contribution ($B_{\rm p}=3.7$~kG, $B_{\rm q}=1.0$~kG). The global field component and distribution of the radial field is similar to those seen in Fig.~\ref{fig:fld_best} but there is no evidence of the small-scale magnetic structure which is needed to fit the linear polarization profiles. We conclude that the Stokes $IV$ analysis of the stellar magnetic topologies is fundamentally limited with respect to the scale of magnetic structures which can be resolved with this method. In the particular case of \cvn\ it can be used to assess the overall, global dipolar-like field topology but, compared to the four Stokes parameter inversion, it cannot provide detailed maps of the distributions of the magnetic field strength and orientation across the stellar surface.

\subsection{Multipolar expansion of magnetic maps}

Spherical harmonic expansion provides a convenient method for quantitative assessment and detailed characterization of the magnetic field maps obtained in the MDI inversions. It also allows us to objectively compare the magnetic field topology of \cvn\ with the field geometry of other Ap stars, in particular \cam\ which was previously studied in all four Stokes parameters by \citet{kochukhov:2004d} using the same inversion technique.

Here we apply the spherical harmonic expansion method introduced by \citet{piskunov:2002a} and employed in the analysis of \cam\ by \citet{kochukhov:2004d}. The three vector components of the magnetic field distribution are approximated with a superposition of real spherical harmonic series, including both poloidal and toroidal expansion terms. In the analysis of \cvn\ we truncate the multipolar expansion at $\ell=10$. This gives us 240 poloidal and toroidal multipolar coefficients, which are determined by solving a linear least-squares problem. Different latitudes on the stellar surface are treated with different weights to account for the variation of the field reconstruction quality and to exclude the invisible part of the stellar surface (hidden to us due to the stellar geometry).

\begin{figure*}[!th]
\centering
\figgps{5.7cm}{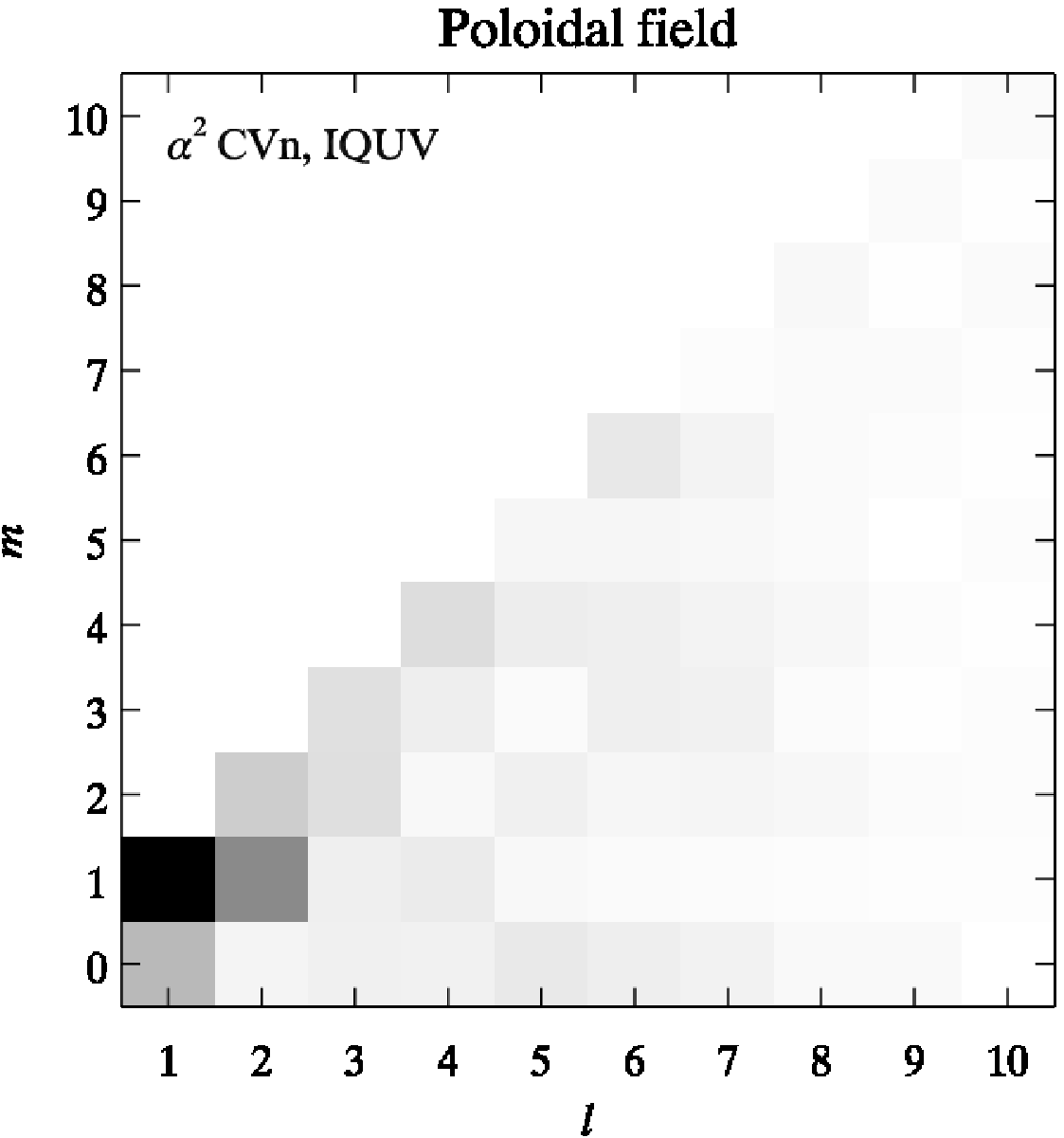}\quad
\figgps{5.7cm}{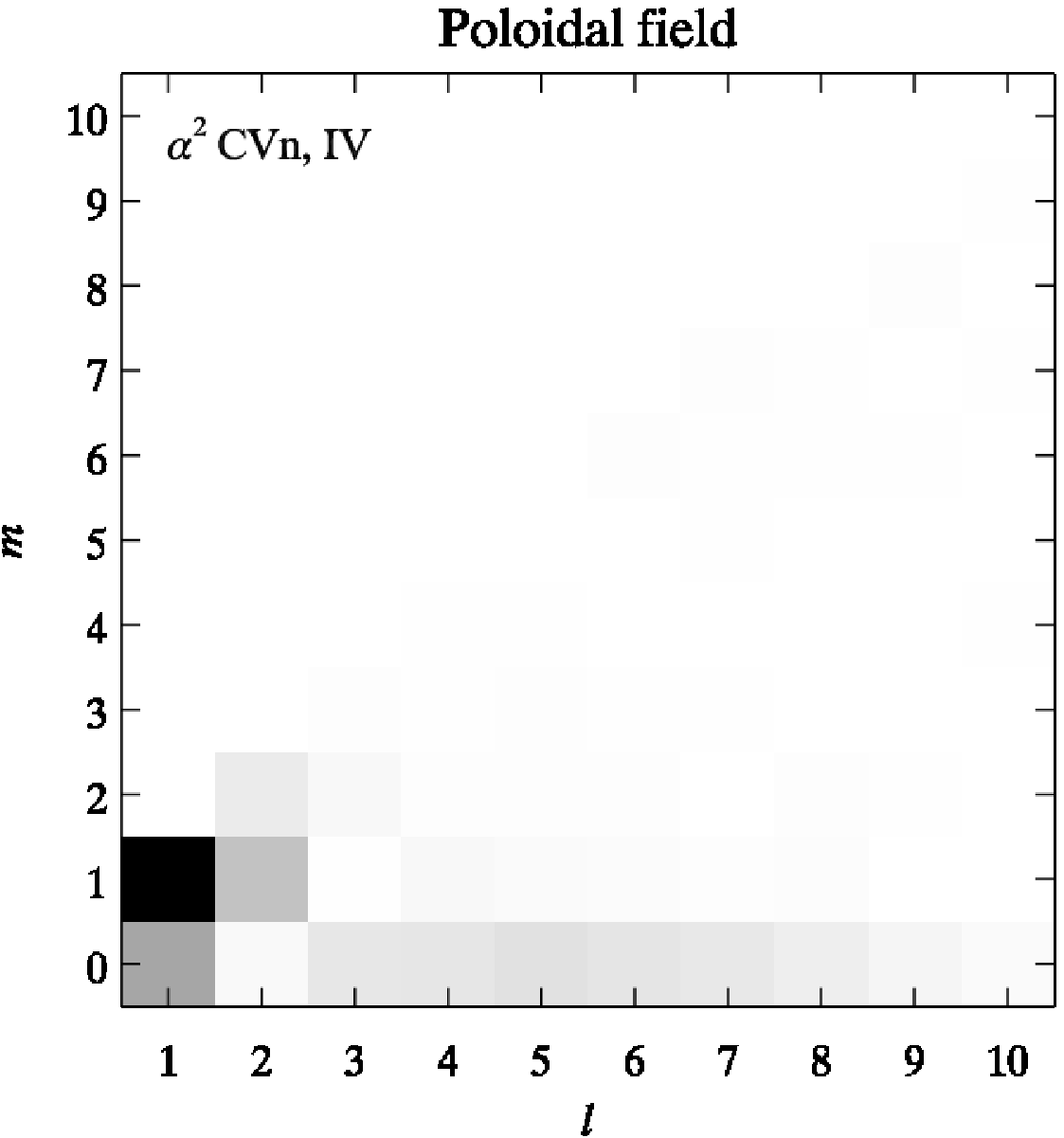}\quad
\figgps{5.7cm}{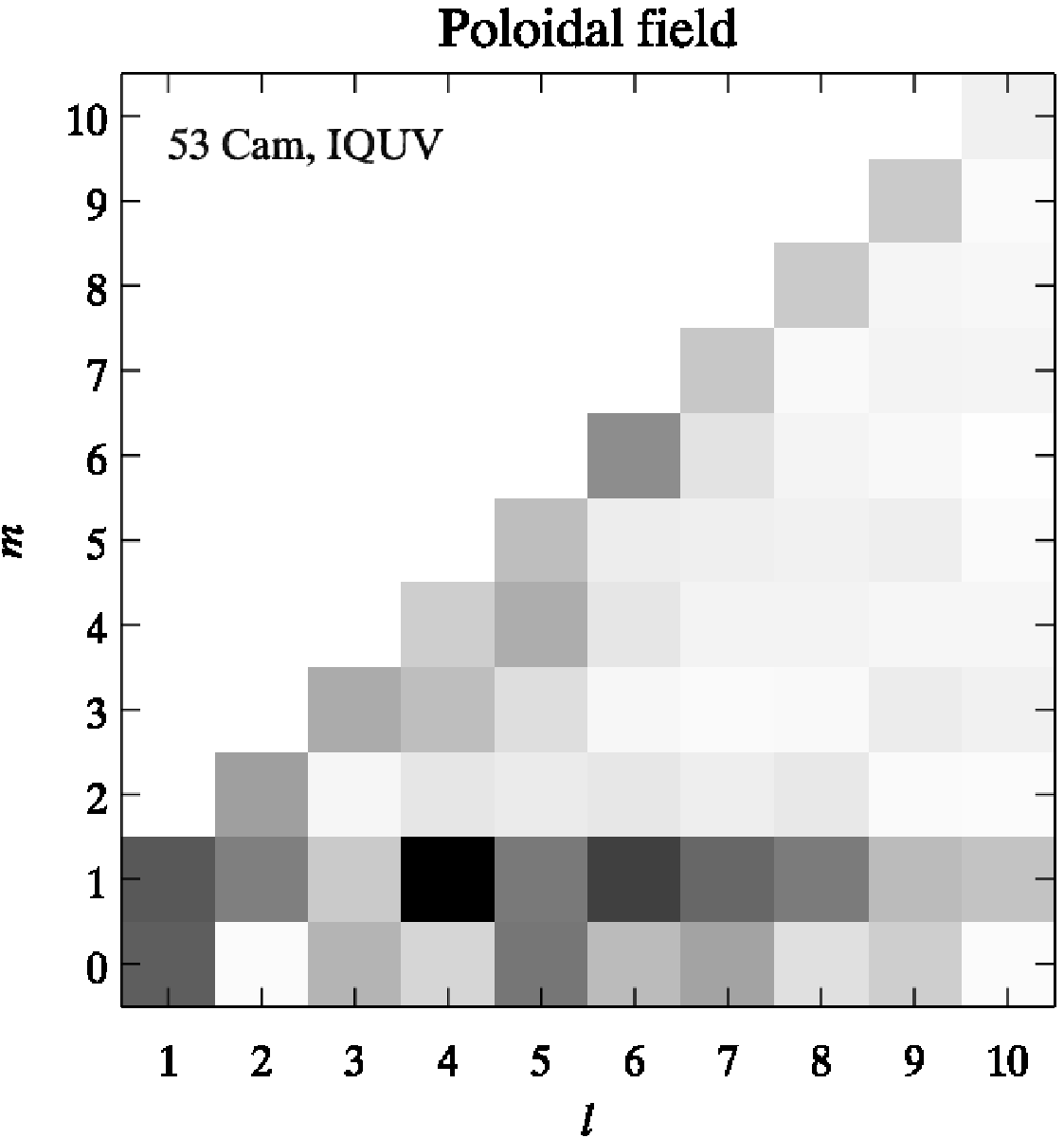}\vspace*{0.3cm}
\figgps{5.7cm}{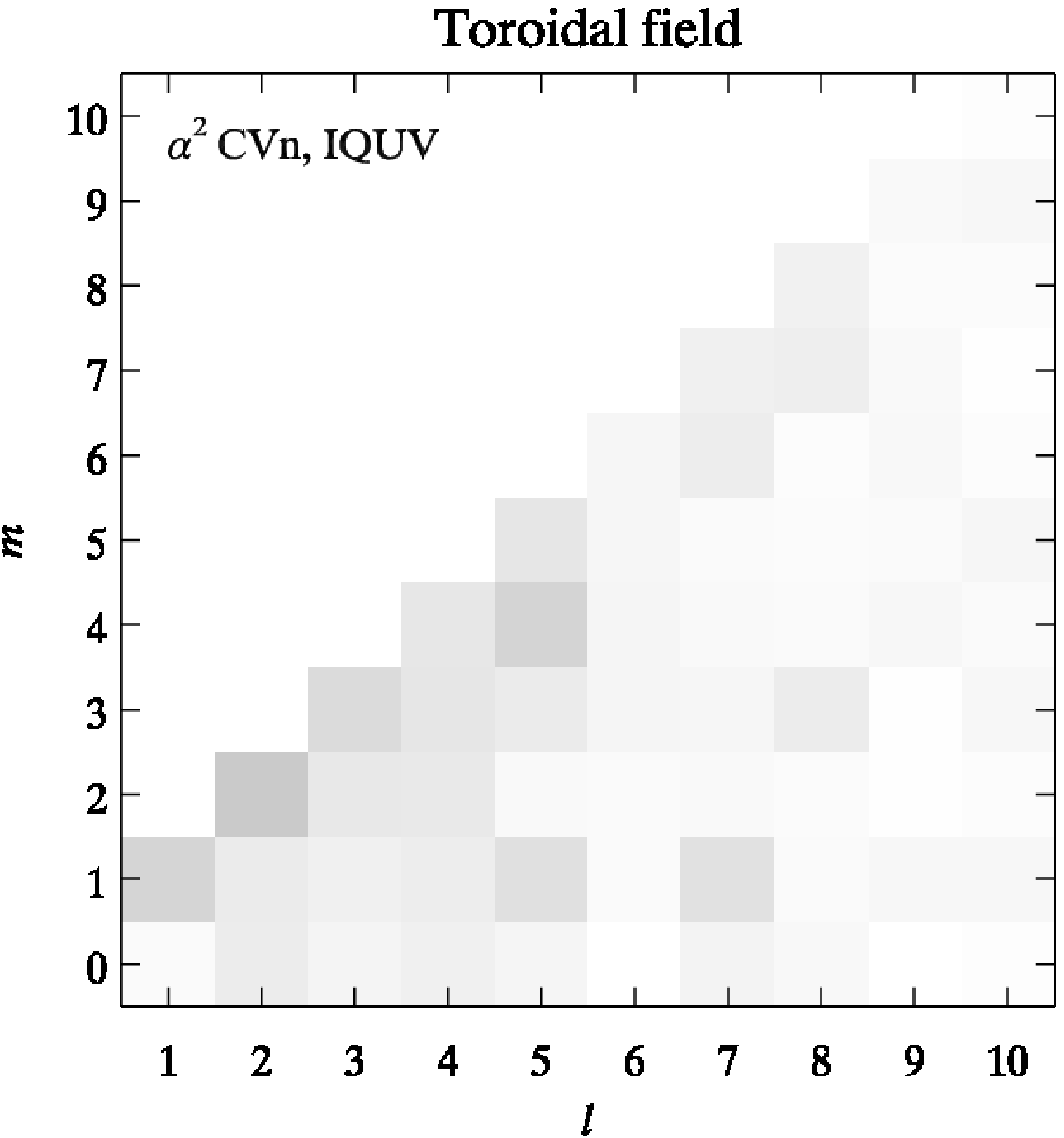}\quad
\figgps{5.7cm}{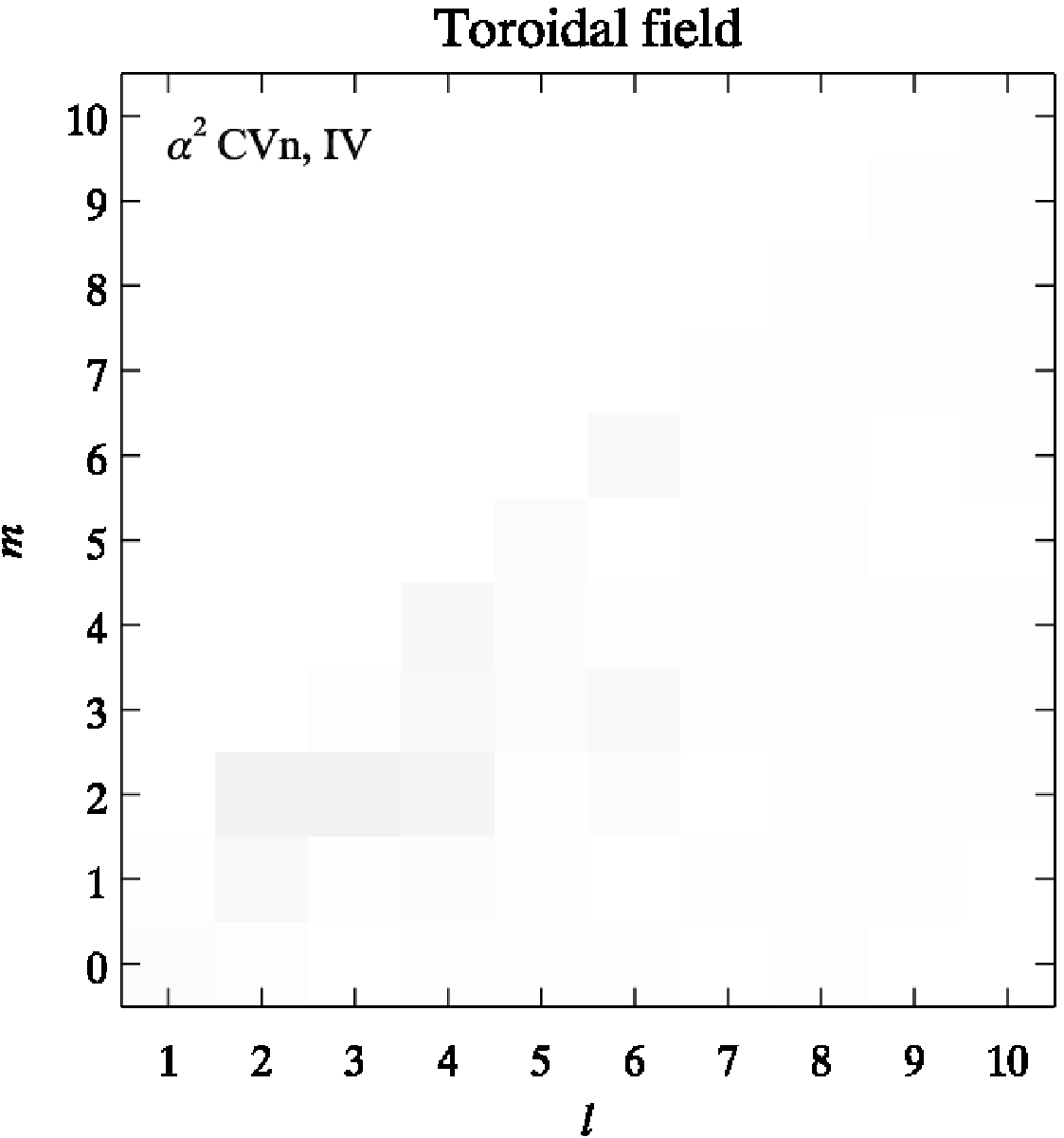}\quad
\figgps{5.7cm}{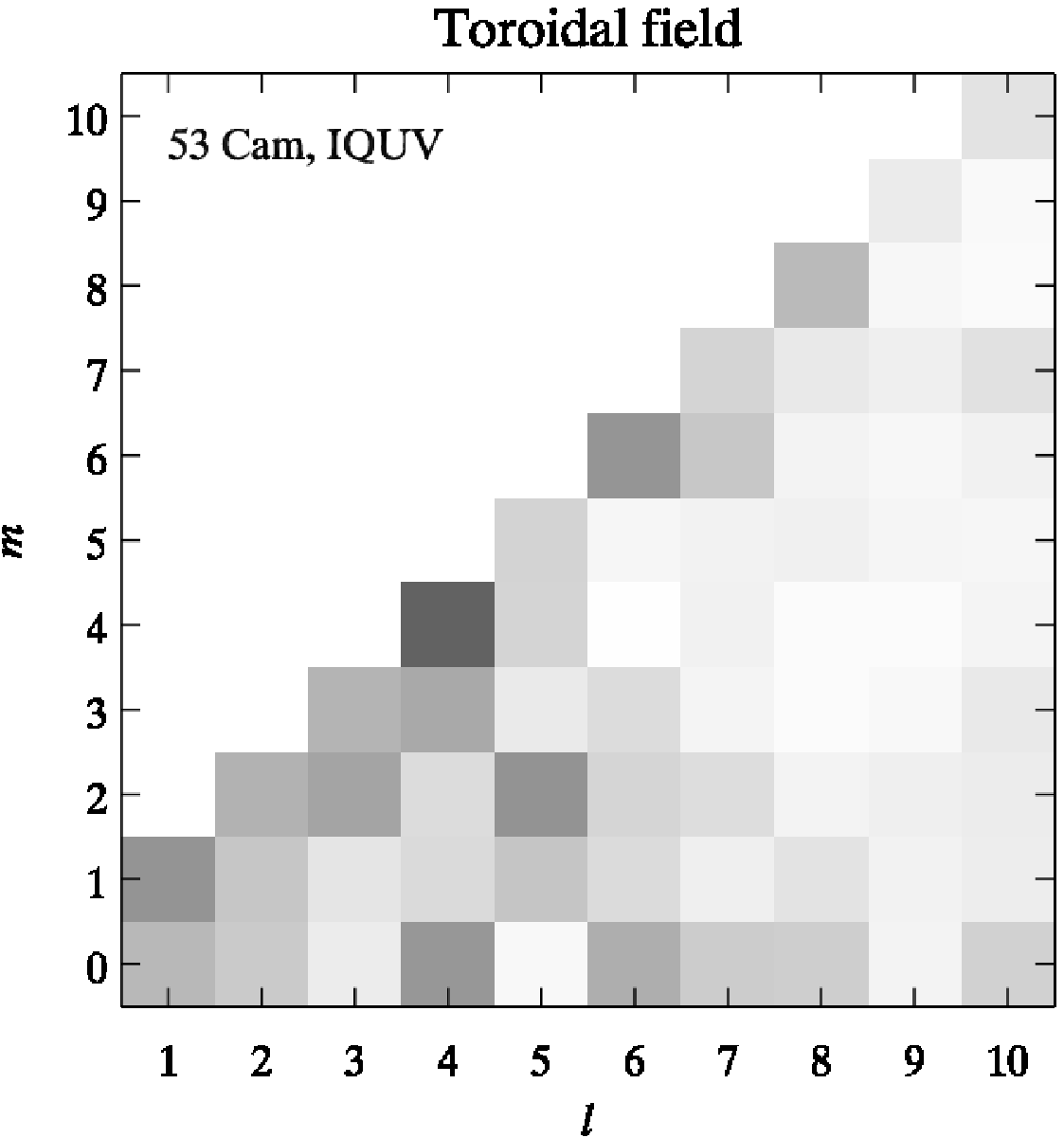}
\caption{Coefficients for the spherical harmonic expansion of the MDI maps of \cvn\ and \cam. The columns correspond to the surface field of \cvn\ inferred from the full Stokes vector inversion (left), magnetic field geometry of \cvn\ obtained from the Stokes $IV$ spectra (middle) and magnetic field of \cam\ \citep{kochukhov:2004d}. The top row shows coefficients for the poloidal field, while the bottom row corresponds to the toroidal field component (practically absent in the MDI maps of \cvn\ reconstructed from the circular polarization spectra).}
\label{fig:expansion}
\end{figure*}

The results of the multipolar expansion are presented in Fig.~\ref{fig:expansion} for the magnetic field topology of \cvn\ obtained from all four Stokes parameters (Fig.~\ref{fig:fld_best}), from the Stokes $IV$ spectra modeled using multipolar regularization (Fig.~\ref{fig:fld_mult}) and for the surface magnetic field of \cam. For the latter star the MDI map was produced with the same modified \inv\ code as applied in the present study of \cvn, using the Stokes $IQUV$ profiles of the three \ion{Fe}{ii} lines studied by \citet{kochukhov:2004d}. The updated magnetic map of \cam\ does not differ appreciably from the average field topology inferred by those authors.

Multipolar expansion coefficients, $\sqrt{(C^m_\ell)^2+(C^{-m}_\ell)^2}$, are plotted in Fig.~\ref{fig:expansion} as a function of $\ell$ and $m$, separately for the poloidal and toroidal expansion terms. The greyscale for each magnetic map is renormalized so that the largest coefficient is black and the smallest is white.
Fig.~\ref{fig:expansion} suggests that the field geometry of \cvn\ is dominated by $\ell=1$ modes. At the same time, the contribution of the higher-$\ell$ modes and the amplitude of the toroidal field components is also non-negligible for the MDI map obtained from all four Stokes parameters (left column in Fig.~\ref{fig:expansion}). For this surface magnetic field distribution we see contributions of modes with $\ell$ up to 5--6. On the other hand, the $IV$ mapping results (middle column in Fig.~\ref{fig:expansion}) suggest a considerably simpler field, showing very clearly the reduction of the information content when linear polarization spectra are excluded from the Doppler Imaging analysis. In this case the dominant dipolar mode is distorted only by a marginal contribution of the non-axisymmetric $\ell=2$ and axisymmetric $\ell=3$--8 components, while the toroidal field is practically absent.

Despite a consistent analysis using the same inversion methodology and \mus\ spectropolarimetric data of similar quality and phase coverage, the level of complexity turns out to be dramatically different for the surface magnetic field topologies of \cvn\ and \cam. For \cvn\ the $\ell=1$ mode is more important relative to other components and the contribution of the toroidal field is small. For \cam\ the power is not concentrated in the dipolar component but spread out over the entire $\ell=1$--10 range, with the broad maximum at $\ell=4$--7, and the toroidal components are noticeably stronger than in \cvn. Therefore, we conclude that the surface magnetic field complexity of the two Ap stars studied using high-resolution four Stokes parameter observations is intrinsically different.

\section{Discussion}
\label{disc}

In this paper we have described what is only the second self-consistent analysis of high-resolution Stokes $IQUV$ line profiles of a magnetic Ap star. For the prototypical spectrum variable \cvn, we have employed phase-resolved polarization spectra to derive a detailed map of the surface magnetic field intensity and orientation using the Magnetic Doppler Imaging approach. In addition, we have derived similarly detailed maps of the surface distributions of the abundances of Fe and Cr. 

We find that the overall {\em structure} of the magnetic field in \cvn\ is dipole-like, with approximately half of the stellar surface covered with the outward-directed radial field while the other exhibits inward-directed field. This is in agreement with the field geometry suggested in previous magnetic analyses of \cvn\ obtained from fitting of multipolar models. However, the field {\em strength} distribution we derive reveals that the field is, in fact, far more complex than the simple geometry suggested by earlier models. In particular, there is a definite asymmetry in the field strength and structure between the negative magnetic pole and the positive pole. The field is clearly stronger at the positive pole and its structure is dominated by high-contrast magnetic spots where the field strength reaches 4.5~kG locally. Interestingly, this is approximately 1~kG higher than the local field at the opposite side of the star. These spots are analogous to those detected in the A4p star \cam\ by \citet{kochukhov:2004c}. We investigated the possibility that the defining characteristic of these maps -- the high-contrast structure of the field strength at smaller spatial scales -- may result from insufficiently smoothing as a consequence of underestimation of the regularization intensity. We thoroughly examined this possibility by conducting inversions with different intensities of the regularization parameter for the magnetic field. We confirmed that our choice of regularization is correct because any substantial increase of the smoothing of the magnetic maps noticeably worsened the agreement between the observed and computed linear polarization observations. We thereby confirmed that the smaller-scale structures in our map are real, and required in order to reproduce the observations. We also computed magnetic field maps using just the Stokes $IV$ spectra. These maps were noticeably less structured than those derived from the full Stokes $IQUV$ data set, confirming that the linear polarization observations are essential to detecting this complex, smaller-scale component of the magnetic field.

A spherical harmonic decomposition was applied to quantitatively study the derived magnetic field topology. This analysis suggests that the field geometry of \cvn\ is dominated by $\ell=1$ modes. At the same time, the contribution of the higher-$\ell$ modes and the amplitude of the toroidal field components is also non-negligible, with contributions of modes with $\ell$ up to 5--6. 

We also re-visited the analysis of \cam\ performed in 2004, obtaining a new field map of this star using the current version of the mapping software. Despite this consistent analysis using the same inversion methodology and spectropolarimetric data of the same origin, and of similar quality and phase coverage, the level of magnetic field complexity is found to be dramatically different for \cvn\ and \cam. For \cvn\ the $\ell=1$ mode is more important relative to other components and the contribution of the toroidal field is small. For \cam\ the power is not concentrated in the dipolar component but is distributed over the entire $\ell=1$--10 range, with a broad maximum at $\ell=4$--7, and toroidal components that noticeably stronger than those of \cvn. Also, the clear hemispheric asymmetry of the qualitative field properties of \cvn\ -- with small-scale structures confined essentially to the stronger, positive pole, the origin of which is not known -- is not obviously reflected in the field of \cam. Therefore, we conclude that the surface magnetic field complexity of the two Ap stars studied using high-resolution four Stokes parameter observations is intrinsically different. The reason for this difference is currently a mystery -- similar maps of a larger sample of stars will be needed before a clearer understanding will be possible.

The surface abundance maps of Cr and Fe show high-contrast distributions of both elements, varying in abundance by 3--4~dex and with pronounced abundance minima at a location corresponding roughly to the negative magnetic pole. We also observed that the map obtained from the two strong \ion{Fe}{ii} lines exhibits noticeably smaller abundances in the region around the negative magnetic pole than those obtained for this star using weaker Fe lines by \citet{kochukhov:2002b}. We discussed how this could be be ascribed to the effects of vertical chemical stratification, which would lead to substantial weakening of the cores of intrinsically strong \ion{Fe}{ii} lines relative to intrinsically weak lines, yielding a smaller abundance if such lines are analyzed neglecting chemical stratification. 

The investigations of the magnetic fields of both \cvn\ (in this paper) and \cam\ (by \citealt{kochukhov:2004c}) lead to the view of a hidden complexity of the magnetic fields of Ap stars -- a complexity which is only revealed with the help of linear polarization measurements. Despite the successes achieved, the data on which these studies have been based is fundamentally limited. The resolving power of the \mus\ spectropolarimeter was rather low -- just 35000 -- and the throughput of the instrument below 1\%. These characteristics led to polarization spectra in which Zeeman signatures were often undetectable (particularly in linear polarization), or only detectable at relatively low significance. The current generation of high-resolution Stokes $IQUV$ spectropolarimeters -- including ESPaDOnS on the Canada-France-Hawaii Telescope, Narval on the T\'elescope Bernard Lyot (both with $R\approx 6.5\times 10^4$ and 20\% throughput) and HARPSpol on the ESO La Silla 3.6m telescope (with $R\approx 1.2\times 10^5$ and 10\% throughput) -- are capable of acquiring data of far greater quality. In particular, observations acquired by \citet{silvester:2008} with ESPaDOnS and NARVAL show the importance of high resolving power for confidently measuring linear polarization. Not only are these instruments capable of acquiring much better data, they also allow us to acquire such data for a much larger sample of stars. Application of these new tools to investigating the detailed magnetic topologies of Ap stars -- with a range of field strengths, rotation rates, ages and masses -- is critical to understanding the physics of their magnetic fields. This includes not only the puzzling structures revealed in the present study, but also the much broader problems of field origin and evolution.

In conclusion we would like to add that our studies of magnetic field geometries of Ap stars unequivocally demonstrate the necessity of using spectra in all four Stokes parameters for reliable reconstruction of the field topologies at various spatial scales. We found that a relatively simple and smooth field structure inferred from circular polarization is invariably superseded by a substantially more complex magnetic topology when linear polarization is incorporated in the Doppler imaging analysis. There are no reasons to believe this trend to be specific to Ap stars. Thus, we suggest that full Stokes vector observations of magnetic stars in other parts of the H-R diagram will lead to a similar dramatic change of the inferred field structure, especially for the late-type active stars which lack a dominant low-order field component. Understanding of stellar magnetism remains fundamentally incomplete without four Stokes parameter spectropolarimetry.

\begin{acknowledgements}
Calculations presented in this paper were carried out at the supercomputer facility provided to the Uppsala Astronomical Observatory by the Knut and Alice Wallenberg Foundation and at the UPPMAX supercomputer centre at Uppsala University. O.K. is a Royal Swedish Academy of Sciences Research Fellow supported by grants from the Knut and Alice Wallenberg Foundation and from the Swedish Research Council. G.A.W.'s research is supported by a Discovery Grant from the Natural Science and Engineering Research Council of Canada, as well as a grant from the Academic Research Program of the Department of National Defence (Canada). 
Resources provided by the electronic databases (VALD, Simbad, NASA ADS) are gratefully acknowledged.
\end{acknowledgements}

%\bibliographystyle{aa}
%\bibliography{astro_papers}

\end{document}